\documentclass[]{article}
\usepackage{amssymb,amsmath}
\usepackage{overcite}
\usepackage{epsf}

\newcommand{\dd}{{\rm d}}
\newcommand{\ee}{{\rm e}}
\newcommand{\eps}{\varepsilon}
\newcommand{\pr}[1]{{\cal P}[#1]}
\newcommand{\rss}{\rho_{0}}

\newcommand{\gss}{g_{0}}
\newcommand{\gpss}{g'_{0}}
\def\d{{\mathrm{d}}}
\newcommand{\erfc}{{\rm erfc}}
\newcommand{\ai}{{\rm Ai}}
\newcommand{\bi}{{\rm Bi}}
\newcommand{\alp}{\left(\frac{2\Gamma}D\right)^{\frac13}}
\newcommand{\blp}{\left(\frac{b\Gamma}{2D}\right)^{\frac12}}
\newcommand{\D}{\displaystyle}
\newcommand{\delt}{\Delta t}
\newcommand{\delx}{\Delta x}              
\newcommand\Or[1]{{\cal O}(#1)}

\begin{document}

\title{
DIFFUSION-LIMITED REACTION IN ONE DIMENSION:
PAIRED AND UNPAIRED NUCLEATION}

\author{Salman Habib\\
Theoretical Division T-8, Los Alamos National Laboratory\\
Los Alamos, New Mexico 87545\\
\and
Katja Lindenberg\\
Department of Chemistry and Biochemistry 0340\\
University of California San Diego\\
La Jolla, California 92093-0340\\          
\and
Grant Lythe\\
Theoretical Division T-8, Los Alamos National Laboratory\\
Los Alamos, New Mexico 87545\\
and\\
GISC, Matem\'aticas, Universidad Carlos III de Madrid\\
Avenida de la Universidad 30,
28911 Legan\'es, Spain\\
\and
Carmen Molina-Par\'{\i}s\\
Theoretical Division T-8, Los Alamos National Laboratory\\
Los Alamos, New Mexico 87545\\
and\\
Centro de Astrobiolog\'{\i}a CSIC-INTA\\
Carretera de Ajalvir, Km. 4\\
28850 Torrej\'on, Madrid, Spain}

\date{March 24, 2001 }
\maketitle


\begin{abstract}
We study the dynamics of diffusing particles in one space
dimension with annihilation on collision and nucleation (creation of
particles) with constant probability per unit time and length. The
cases of nucleation of single particles and nucleation in pairs are
considered.  A new method of analysis permits exact calculation of the
steady-state density and its time evolution in terms of the three
parameters describing the microscopic dynamics: the nucleation rate,
the initial separation of nucleated pairs, and the diffusivity of a
particle.  For paired nucleation at sufficiently small initial
separation the nucleation rate is proportional to the square of the
steady-state density. For unpaired nucleation, and for paired
nucleation at sufficiently large initial separation, the nucleation
rate is proportional to the cube of the steady-state density.
\end{abstract}

\section{Introduction}
\label{intro}

Reaction rates controlled by collisions between diffusing particles
depend on the distribution of distances between particles as well as
on the density of particles.  In particular, as Noyes stated in 1961
``{\em Any rigorous treatment of chemical kinetics in solution must
consider concentration gradients that are established by the existence
of the reaction itself}".~\cite{noyesrev} Here, we study the dynamics of
point particles in one dimension, nucleated at random positions and
times, then diffusing until colliding with and annihilating another
particle. Competition between nucleation and annihilation produces a
statistically steady state with a well-defined mean density of
particles and distribution of distances between particles.  We shall
contrast two types of nucleation: {\em unpaired}, in which particles
are deposited at random locations at random times, and {\em paired},
in which {\em pairs} of particles are deposited at random locations.
The dynamics is as follows: \begin{enumerate} \item[(i)] Particles are
nucleated in pairs with initial separation $b$;

\item[(ii)]
Nucleation occurs at random times and positions with rate $\Gamma$;

\item[(iii)]
Once born, all particles diffuse independently with diffusivity $D$;
and

\item[(iv)]
Particles annihilate on collision.
\end{enumerate}

\begin{figure}
\begin{center}
\leavevmode
\epsfxsize = 4.0in
\epsffile{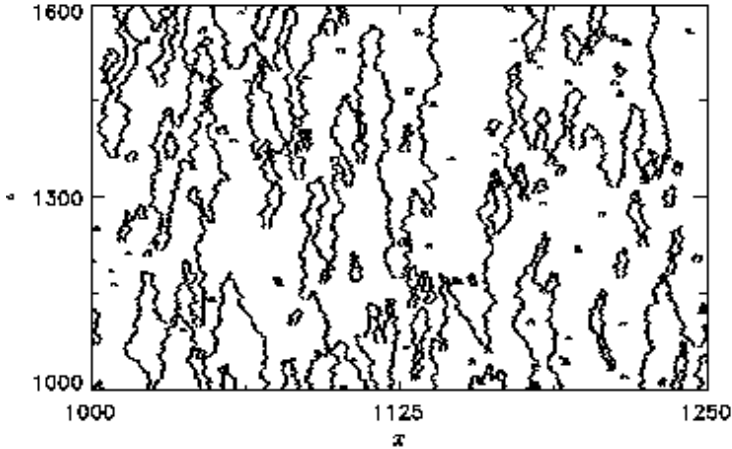}     
\end{center}
\caption{
One part of a numerical solution with paired nucleation, 
governed by (i)-(iv).
Time increases upwards and each dot indicates the space-time
position of a diffusing particle.
\protect\(\Gamma = 1.25\times10^{-3}\), \(b = 2\), \(D = 0.1\)
}
\label{spacetime}
\end{figure}

A portion of a typical realization of these dynamics is shown in
Fig.~\ref{spacetime}.
For unpaired nucleation (i) and (ii) are replaced by
\begin{enumerate}
\item[(i')]
Particles are nucleated at random times and positions with rate $Q$.
\end{enumerate}

An existing method of analysis, based on a truncated hierarchy of
correlation functions, is developed and extended in this article to
the case of paired nucleation, yielding expressions for the
correlation functions in the steady state, and for the time scales for
relaxation towards the steady state. We also introduce a different method of
analysis that yields an {\em exact} explicit expression for the steady-state
density and for the time dependence of the density starting from
arbitrary initial conditions. Our analytical predictions are compared
with the results of direct numerical simulations. In the simulations,
large numbers of diffusing particles are simultaneously evolved in
continuous space, with
annihilation whenever two paths cross and nucleation (paired or
unpaired) at random times and positions.

A striking difference between paired and unpaired nucleation
is the scaling of the steady state density of particles, \(\rss\), with
the nucleation rate: \(\rss \propto \Gamma^{\frac12}\) (paired)
versus \(\rss \propto Q^{\frac13}\) (unpaired). Here, we
shall exhibit the crossover between these two cases
in terms of the following dimensionless quantity:
\begin{equation}
   \eps = \left(\frac{2\Gamma}D\right)^{\frac13}\,b.
\label{epsdef}
\end{equation}
For \(\eps\to\infty\), the dynamics described by (i)--(iv)
is equivalent to that described by (i'), (iii)--(iv) with
the replacement
\begin{equation}
   Q \to 2 \Gamma.
\label{qto2g}
\end{equation}

The paper is arranged as follows. In the remainder of this section we
summarize published results for reaction-diffusion systems.  In
Sec.~\ref{vk} we analyze the dynamics using a hierarchy of
equations for particle density functions, called ``reduced
distribution functions'' by Van Kampen~\cite{vankampen}.  Derivation
of the reaction kernel leads to an exact relation between the density
of particles and the derivative of the correlation function.
We also explore the linear response to a perturbation away
from the steady state to establish the time scales for relaxation. In
Sec.~\ref{exact}, by introducing a function that satisfies a closed
linear partial differential equation, we present exact expressions for
the steady-state density and for the time evolution of the density
with arbitrary initial conditions. In particular, analytical results
are presented describing the rapid initial annihilation that
transforms an initially random distribution into one characterized by
an effective repulsion between particles.

\subsection{Unpaired nucleation}

Analysis of diffusion-limited reaction dates back to
M. von Smoluchowski. His {\em Mathematische Theorie
der raschen Koagulation}~\cite{vonsmol} considered reaction
between diffusing particles resulting in merger, with the reaction
taken to occur immediately whenever two particles are a distance \(R\)
apart. He introduced a diffusion equation for the density of particles
relative to the position of a test particle and noted that the density
is zero at all times at radius \(R\) \cite{heir}.
For many years it was assumed that the final result of a complete 
calculation following the procedure outlined by Smoluchowski would be
an equation for the mean density of particles, \(\rho\), of the
form~\cite{vankampen}
\begin{equation}
   \dot \rho = Q - k_s\rho^2,
\label{naive}
\end{equation}
where \(Q\) is the rate (per unit length and time) of appearance of
new particles and \(k_s\) is constant. This would imply, for the
 case {\em without nucleation} (\(Q=0\)), that the density is
proportional to \(t^{-1}\) for \(t\to\infty\). However,
 arguments based on dimensional analysis and scaling show that 
this is not true in one dimension~\cite{oandz,degennes,tandw,nandk,kandr}.
In 1983, Torney and McConnell studied this case and published an
exact solution for the mean density as a function of
time~\cite{tandm}. Starting from an initial random distribution of
particles, they found
\begin{equation}
   \rho(t) = \rho(0)\exp\left(8Dt\rho^2(0)\right)
   \erfc\left(\rho(0)(8Dt)^{\frac12}\right).
\label{tandmdens}
\end{equation}
In particular, \(\rho \to (8\pi D t)^{-\frac12}\) for \(t\to\infty\).
A rederivation of the result of Torney and McConnell was provided by
Spouge~\cite{spouge}, whose insight was that an annihilation process
is equivalent to a coagulation process if coagulants made up of an
even number of particles are considered as diffusing
``ghosts.'' Derivations based on reflection principle~\cite{balding}
and field theory~\cite{field,lushnikov} methods have also been
published.

In discrete models of diffusion-limited reaction, diffusion is
approximated by hopping between neighboring sites on a lattice. Here,
too, the density of particles without nucleation is proportional to
\(t^{-\frac12}\) for \(t\to\infty\)
\cite{lushnikov,bandg,bandl,privman,kpwh}.  Moreover, with unpaired
nucleation, the steady state density is proportional to the third
power of the nucleation rate~\cite{racz,da,abd}. This can be
interpreted as evidence for a time-dependent rate constant \(k_s\) in
\eqref{naive}, or as requiring \eqref{naive} to be replaced by an
equation of the form
\begin{equation}
   \dot \rho = Q - k_c\rho^3.
\label{cubic}
\end{equation}
However, no polynomial equation for the density can
describe both the steady state with nucleation and the long-time decay
of the density without nucleation~\cite{da,lin}.

An exact solution has been found in one dimension for a discrete
coagulation model with one fixed source.  The latter solution is
related to the probability that a given spin in an Ising chain with
random initial conditions does not change its value before time $t$
\cite{dhp}.  For discrete and continuous coagulation models, exact
results are available not only for the density but also for the
spectrum of relaxation rates,~\cite{da} the distribution of
interparticle distances,~\cite{da} and correlation
functions.~\cite{bena}  They are obtained by
considering the function \(E(n\Delta x,t)\), defined as the
probability that an arbitrarily chosen segment of \(n\) consecutive
sites contains no particles, satisfying a closed kinetic equation. It
has, however, not proven possible to extend this method to the case of
annihilation on contact, because the function \(E(n\Delta x,t)\) does not
satisfy a closed equation~\cite{abd}.

\subsection{Paired nucleation}

The ``coefficient of recombination'' of two particles initially close
together was introduced in the study of subatomic particles
\cite{langevin}.  The relative motion of two diffusing particles is
equivalent to a problem of Brownian motion of one
particle.~\cite{onsager,expstep}

A discrete model that corresponds to paired
nucleation is the Ising model, with nucleation at neighboring
sites. Its dynamics was studied analytically by Glauber in 1963
\cite{glauber}; the nucleation rate is proportional to the square of
the steady state density for nucleation rates sufficiently small that
excluded volume effects can be neglected~\cite{racz}. Computer
simulations of a discretized reaction-diffusion model \(A+B\to 0\),
published in 1987 \cite{aandk}, contrasted the scalings of the steady
state density according to whether nucleation occurred at random sites
or in pairs at neighboring sites. In the latter case,
the scaling \(\Gamma \propto \rho^2\) was found.

A different approach to diffusion-limited reaction was recently
introduced in the context of kink dynamics in a stochastic
partial differential equation (PDE)~\cite{nucl}. There, the dynamics
was termed ``mesoscopic'' because
it was an approximate model that ignored the internal structure of
kinks and antikinks, treating them simply as particles that happen to
be nucleated in pairs. The treatment was based on classifying
particles according to whether they are annihilated in a collision
with their nucleation partner (recombination) or with a different
particle (nonrecombinant annihilation). 
The steady-state density \(\rss\) is related to the mean lifetime of
a particle, \(\tau\), by 
\begin{equation}
   \rss = 2\Gamma\tau.
\label{taudef}
\end{equation}
The mean lifetime \(\tau\) was estimated directly
by averaging over the possible histories of a
pair of particles born together. This approximate analysis
yielded the estimate \(\rss = (3b\Gamma/8D)^{\frac12}\).

\section{Hierarchy of distribution functions}
\label{vk}

Let \(f_n(x_1, \dots x_n;t)\d x_1\ldots\d x_n\) be the probability
that there is one particle in \((x_1,x_1+\d x_1)\), one in
\((x_2,x_2+\d x_2)\), $\dots$, and one in \((x_n,x_n+\d x_n)\) at time
\(t\), regardless of the positions of the other
particles~\cite{vankampen}.  The function \(f_1(x_1;t)\) is
the particle density at \(x_1\) at time \(t\). On deriving the differential
equation for its time derivative,
one finds that it involves \(f_2(x_1,x_2 ;t)\)~\cite{vankampen,heir,lin,kak}.
Similarly, the time derivative of \(f_2(x_1,x_2 ; t)\) involves
\(f_3(x_1,x_2,x_3; t)\). One is thus led to a hierarchy of differential
equations for the evolution of the distribution functions.

In this section we derive the source terms appropriate for paired
nucleation in the hierarchy of differential equations.  We also derive
the reaction terms corresponding to diffusion with annihilation on
collision, without needing to introduce a reaction radius. Three
parameters remain in the theory for paired nucleation: the nucleation
rate of pairs \(\Gamma\), their separation at nucleation \(b\) and the
diffusivity of a particle \(D\).  For unpaired nucleation there are
two parameters: the nucleation rate \(Q\) and the diffusivity of a
particle \(D\). The annihilation process is immediate on collision and
therefore does not require extra parameters. It manifests itself
instead in boundary conditions on the distribution functions.  We shall
truncate the hierarchy of distribution functions using an ansatz
for the three-point correlation function, introduced in the literature
for unpaired nucleation~\cite{lin,kak}, thus obtaining a closed pair
of differential equations for the density and two-point correlation
function.  Their solution yields analytical approximations for the
steady-state density and two-point correlation function.
By examining perturbations
away from the steady state, we derive the time scales for
relaxation towards the steady state.

The evolution  of the reduced distribution functions has a number
of contributions
\begin{eqnarray}
   \frac{\partial}{\partial t} f_n (x_1, \dots x_n;t)&=&
   D\nabla^2 f_n(x_1, \dots x_n;t)
\nonumber\\ [12pt]
&&   - \sum_{(ij)}^{n} k(x_i ,  x_j)
   f_n(x_1, \dots x_n;t)\nonumber\\ [12pt]
   &&- \sum_{i=1}^{n} \int_{-L}^L \d x_{n+1} \ k(x_i , x_{n+1} )
   f_{n+1}(x_1, \dots x_n, x_{n+1};t)\nonumber\\  [12pt]
   && + \ \textrm{sources}.
\label{hierarchy}
\end{eqnarray}
First on the right is the diffusion term, due to
the motion of each particle with diffusion coefficient $D$.
  The second term represents the reaction
between two of the $n$ particles:
$k(x,x')$ is the probability per unit time that a particle at $x$ and one
at $x'$ react, and the summation
is over all pairs that can be selected from the $n$
particles.  The third term accounts for the fact that
each of the $n$ particles may react with another that is not part of
the set of $n$ particles, and $2L$ is the size of the
system.  The last term is a source contribution
whose form is given in detail below.  Equation~\eqref{hierarchy} is one
in an infinite hierarchy.
Explicitly, the first two equations in the hierarchy are:
\begin{eqnarray}
  \frac{ \partial}{\partial t} f_1(x_1;t)&=& D
  \frac{ \partial^2}{\partial x_1^2} f_1(x_1;t)
   - \int_{-L}^{L} \d x'\; k(x_1,x') f_2 (x_1, x';t)+q_1(x_1),
   \label{f1dot} \nonumber\\ 
   && \\ [12pt]
  \frac{ \partial}{\partial t} f_2(x_1,x_2;t)&=& D\nabla^2
   f_2(x_1, x_2;t)
   - k(x_1,x_2) f_2( x_1, x_2;t) \nonumber\\ [12pt]
   &&- \int_{-L}^L \d x'[ k( x_1,x') + k( x_2,x') ]
   f_3 (x_1 , x_2, x';t)\nonumber\\ [12pt]
   && +\ q_2(x_1,x_2) + f_1(x_1;t)q_1(x_2) + f_1(x_2;t)q_1(x_1).
      \label{f2dot}
\end{eqnarray}

\subsection{Source terms for paired nucleation}

The term $q_1(x_1)$ in \eqref{f1dot} and \eqref{f2dot} is the
probability density per unit time for the creation of a particle at
\(x_1\); the term $q_2(x_1,x_2)$ is the probability
density per unit time for the simultaneous creation of a particle at
$x_1$ and another at $x_2$. When creation of particles always occurs
in pairs, these two source functions are related 
\begin{equation}
   q_1(x_1) = \int_{-L}^L \d x_2 \ q_2(x_1,x_2).
\end{equation}
When the particle creation rates are independent
of position and time, \(q_1(x)\) is constant
\begin{equation}
   q_1(x) = 2\Gamma,
\label{q1}
\end{equation}
and \(q_2(x,x+y)\) is independent of \(x\).
The constant \(\Gamma\) is the rate of creation of pairs per unit length.

Here, because particles are indistinguishable, \(q_2(x_1,x_2)\)
depends only on \(y=|x_1-x_2|\) and the functions
\(f_n(x_1,\ldots,x_i,\ldots,x_n;t)\) are independent of the order of
the \(x_i\).
The probability that two particles initially at $x_1$ and $x_2$ react is
the probability that they diffuse and collide.
Since particle diffusion is isotropic and independent of
position, \(k(x_1,x_2)\) also depends only on \(y=|x_1-x_2|\). We
therefore define 
\begin{equation}
   q(y)\equiv q_2(x_1,x_2) \quad\textrm{ and } \quad
   K(y) \equiv k(x_1,x_2).
   \label{kdef}
\end{equation}
The function \(q(y)\) describes the probability
density of distances between particles nucleated simultaneously.
We shall use the following forms for this function, 
corresponding to unpaired nucleation and to paired nucleation
with initial separation \(b\):
\begin{eqnarray}
   q(y) = \left\{\begin{array}{lll}\Gamma/L
            \qquad &\textrm{unpaired};\\[10pt]
            \Gamma\delta(y-b)
            \qquad &\textrm{paired}.
            \end{array}\right.
\label{q2}
\end{eqnarray}

We can now rewrite \eqref{f1dot} and \eqref{f2dot} as follows:
\begin{eqnarray}
  \frac{ \partial}{\partial t} f_1(x;t)&=& 
  D\frac{ \partial^2}{\partial x^2} f_1(x;t)
   - 2\int_{0}^{L} \d y \; K(y) f_2 (x,x+y;t) + 2\Gamma,
   \label{f1dota}\nonumber\\&&\\ [12pt]
  \frac{ \partial}{\partial t} f_2(x,x+y;t)&=& 
   D\nabla^2 f_2(x, x+y;t)
   - K(y) f_2( x,x+y;t) \nonumber\\ [12pt]
   &&- \int_{-L}^L \d z\left( K( |z| ) + K( |z-y| ) \right)
   f_3 (x , x+y , x+z ;t)\nonumber\\ [12pt]
   && +\ q(y) + 2\Gamma(f_1(x;t) + f_1(x+y;t)).
      \label{f2dota}
\end{eqnarray}

If the initial conditions are homogeneous, then the
functions \(f_1\), \(f_2\), \(\ldots\) will be homogeneous at
all times.  In particular, \(f_1(x;t)\) will be independent of \(x\) 
at every \(t\). Let
\begin{eqnarray}
   \rho(t)&\equiv& f_1(x;t),\nonumber\\ [10pt]
   F_n(x_2,\ldots,x_n;t)&\equiv& 
   {\rho^{-n}(t)}{f_n(x,x_2-x,\ldots,x_n-x;t)}.
\label{gdef}
\end{eqnarray}
We shall in particular be interested in the dimensionless correlation
function defined by
\begin{equation}
   g(y,t) \equiv F_2(y;t).
\label{gydef}
\end{equation}
The function \(g(y,t)\) is the probability density at time \(t\) of
particles at a distance \(y\) from a reference particle,
divided by the overall density of particles. It is constructed
numerically as follows. Choose a sample of \(N\) reference particles,
located at \(\{x_i, i=1,\ldots , N\}\) at time \(t\).
For each \(x_i\), construct 
\[G_i(y,t)= \{\textrm{number of particles between}
\ x_i\ \textrm{and}\ x_i+y\}\] for \(y>0\). 
Then \(G(y,t)\) is the average over the 
\(N\) particles of the \(G_i(y,t)\) and
\begin{equation}
   g(y,t) = (\rho(t))^{-1}\frac{\partial}{\partial y}G(y,t).
\label{gycon}
\end{equation}
If there is no correlation between particles at time \(t\),
then \(g(y,t)\)=1 for all $y \geq 0$. In all the situations considered
here, the total length \(2L\) of the system is sufficently
large compared to the correlation length so that \begin{equation}
   \lim_{y\to \infty}\gss(y) = 1,
\label{glim}
\end{equation}
where \(\gss (y)\) denotes the steady state correlation function.

In terms of \(\rho(t)\) and \(g(y,t)\), Eqs.~\eqref{f1dota}
and \eqref{f2dota} now simplify to the pair of equations
\begin{eqnarray}
   \dot\rho(t) &=& -2 \rho^2(t) \int_{0}^L \d y\; \ K(y)g(y,t) + 2\Gamma,
   \label{rhodot1}\\ [12pt]
   \frac{\partial}{\partial t}
   g(y,t)&=&2D\frac{\partial^2}{\partial y^2}g(y,t) - K(y)g(y,t) 
   \nonumber\\ [12pt] && -\
   {\rho(t)}\int_{-L}^L\d z\left(K(|z|)+K(|z-y|)\right)F_3(y,z;t)
   \nonumber\\ [12pt] && 
   +\ 4\Gamma{\rho^{-1}(t)} +\ {\rho^{-2}(t)}{q(y)} 
   -\ 2\rho^{-1}(t)g(y,t)\dot \rho(t).
\label{gdot1}
\end{eqnarray}

\subsection{The reaction kernel}

To complete the description of the dynamics of the system, we consider
the reaction terms for the case where {\em particles diffuse with
diffusivity \(D\) and annihilate on collision}.  We shall see that
a consequence of annihilation on collision is that \(g(0)=0\)
for all $t>0$, where \(g(y)\) is the correlation function defined in
\eqref{gydef}. We derive an exact relation between \(g'(0)\) and the
rate of collisions between particles.

Let \(s(y,\Delta t)\) be the probability that two particles, with
initial separation \(y\), collide before \(\Delta t\). Then
the reaction kernel \(K(y)\), defined in \eqref{kdef}, is given by
\begin{equation}
   K(y) = \lim_{\Delta t \to 0} \frac1{\Delta t}s(y,\Delta t).
\label{kr}
\end{equation}
If both particles diffuse with diffusivity \(D\) then~\cite{expstep,kands}
\begin{equation}
   s(y,\Delta t) = \erfc
   \left(\frac{y}{(8D\Delta t)^{\frac12}}\right),
\label{kerdef}
\end{equation}
where we assume \(L \gg (D\Delta t)^{\frac12}\). 
To calculate the frequency of collisions between particles, we consider
a time interval $t, t+\Delta t$.  Given the density $\rho(t)$ and the
correlation function $g(y,t)$ defined in \eqref{gydef}, we can imagine
following the paths of all the particles from time $t$ to time $t+\Delta
t$ without removing those that collide.  Then, the probability that a
particle chosen at random undergoes a collision between time $t$ and
time $t+\Delta t$ is $P(t, \Delta t)$, where
\begin{equation}
   P(t,\Delta t) = 2\rho(t) \int_0^L \d y \ s(y,\Delta t)g(y,t).
\label{pdef}
\end{equation}
The expression \eqref{pdef} overestimates the number of collisions in
the system with annihilation on collision due to the possibility that
the same particle undergoes two (or more) collisions in the interval
$t, t+\Delta t$.  However, this latter probability is proportional to
$(\Delta t)^2$ as $\Delta t \to 0$, and so \eqref{pdef} is valid for our
system in the limit $\Delta t \to 0$.

Next, consider the dynamics of the system as a whole.  The mean number
of distinct collisions between time \(t\) and time \(t + \Delta t\) is given 
by \(L\rho(t)P(t,\Delta t)\). We can, therefore, write
\begin{eqnarray}
   && \int_0^L \d y\,s(y,\Delta t) g(y,t) \nonumber \\[12pt]
   \qquad & = &
   \int_0^L \d y\,s(y,\Delta t)
   \left(g(0,t) + yg'(0^+,t) + \frac12y^2g''(0^+,t) + \ldots\right)
   \nonumber \\[12pt]
   &=& 
   \left(\frac1{\surd{\pi}}(8D\Delta t)^{\frac12}g(0,t)
   + \frac14(8D\Delta t)g'(0^+,t) + {\cal O}(D\Delta t)^{\frac32}\right).
   \nonumber\\
\label{kercalc}
\end{eqnarray}
The number of collisions between time \(t\) and 
\(t + \Delta t\) is proportional to \(\Delta t\) if
\begin{equation}
   g(0,t) = 0.
\label{gcond}
\end{equation}
Because the number of nucleation events between time \(t\) and time
\(t + \Delta t\) is proportional to \(2L\Delta t\), the condition
\eqref{gcond} is necessary if there is to be a steady-state balance
between nucleation and annihilation. More generally, it is necessary
if \(\rho(t)\) is to obey a differential equation.  It is, of course,
possible to construct initial conditions that do not satisfy
\eqref{gcond}: a random distribution of particles, for example. Then
the number of annihilation events will initially be proportional to
\(t^{\frac12}\); this period of rapid annihilation creates a
``depletion zone''~\cite{noyesrev,da,noyes,csk} in \(g(y,t)\), which
thereafter satisfies \eqref{gcond}.  That \(g(y,t)<1\) for \(y \to
0\) implies an effective repulsion: particles are {\em less} likely
to be found close to a reference particle than a large distance from it.

Using \eqref{kercalc} and \eqref{gcond} gives
exact expressions for the evolution of the density:
\begin{equation}
\left\{\begin{array}{lll}
   \dot\rho(t) &=& -4 D \rho^2(t) g'(0^+,t) + 2\Gamma,
            \qquad \textrm{paired};\\[12pt]
   \dot\rho(t) &=& -4 D \rho^2(t) g'(0^+,t) + Q,
            \qquad \textrm{unpaired}.
            \end{array}\right.
\label{ddens}
\end{equation}
In particular, we have the following relationship
between the steady state density and the derivative of 
the correlation function.
Let \(\rss\) and \(\gss(y)\) denote the steady state density and
correlation function. Then
\begin{equation}
\left\{\begin{array}{lll}
   \Gamma = 2D\rss^2 \gpss(0^+)
            \qquad \textrm{paired};\\[12pt]
   Q = 4D\rss^2 \gpss(0^+)
            \qquad \textrm{unpaired}.
            \end{array}\right.
\label{ssdens}
\end{equation}

The reaction kernel \(K(y)\) is a singular function
\begin{equation}
   \int_0^L \d y \ K(y)g(y,t)  = 
   \D\lim_{\Delta t \to 0} \frac1{\Delta t}
   \int_0^L \d y \ s(y,\Delta t)g(y,t)  = 2D g'(0^+,t).
\label{keval}
\end{equation}
We have assumed that \eqref{gcond} holds.
The derivative of \(g(y)\) is one-sided
\begin{equation}
   g'(0^+,t) \equiv \lim_{a\to 0^+} \frac{g(a,t)}{a},
\label{gzddef}
\end{equation}
because \(g(y)\) is only defined for \(y>0\). In other words
\begin{equation}
   K(y) =  2D \lim_{a\to 0^+} \frac{\delta(y-a)}{a}.
\label{kereval}
\end{equation}
It is interesting to compare \eqref{kereval} with the form of
\(K(y)\) used, for example, by Lindenberg {\em et al.}~\cite{kak}
\begin{equation}
   K(y) = k\delta(y-a),
\label{kerapp}
\end{equation}
which introduced a reaction radius \(a\) and a rate coefficient $k$.
There it was assumed that these constants are connected to the
diffusivity via the Smoluchowski relation $k=2D/a$ and that in the
limit $a\to 0$, $k\to\infty$ their product remains finite.
In the form used here, by contrast, we are able to explicitly take the
limit of zero reaction radius: \(a\to 0\). 

Similarly to \eqref{keval}
\begin{equation}
   \int_{-L}^L \d z \;  K(|z|)F_3(y,z;t) = 4DF_3'(y,0^+;t),
\label{f3eval}
\end{equation}
where
\begin{equation}
   F_3'(y,0^+;t) = \lim_{a \to 0^+} a^{-1}F_3(y,a;t),
\label{f3dderf}
\end{equation}
and
\begin{equation}
   \int_{-L}^L \d z  \; K(|z-y|)F_3(y,z;t) = 4DF_3'(y,y^+;t).
\label{f3evalbis}
\end{equation}
A similar expression was derived for a discrete coagulation
model without nucleation by Lin, Doering, and ben-Avraham~\cite{lin}.
Since \(F_3(y,0;t) = F_3(y,y;t)\), Eqs.~\eqref{rhodot1} 
and \eqref{gdot1} simplify to the pair of equations
\begin{eqnarray}
   \dot\rho(t) &=& -4 D \rho^2(t) g'(0^+,t) + 2\Gamma,
   \label{rhodot2}\\ [7pt]
   \frac{\partial}{\partial t}
   g(y,t)&=&2D\frac{\partial^2}{\partial y^2}g(y,t) 
   -K(y)g(y,t)
   -\ 8D \rho(t)F_3'(y,0^+;t)    \nonumber\\ [4pt] && 
   +\ 4\Gamma{\rho^{-1}(t)} +\ {\rho^{-2}(t)}{q(y)}
   -\ 2\rho^{-1}(t)g(y,t)\dot \rho(t).
\label{gdot2}
\end{eqnarray}
Annihilation on collision is described by the terms involving the
reaction kernel \(K(y)\).
 
\subsection{Truncation of the hierarchy}

We have obtained exact expressions for the evolution of the
density. However, to obtain a closed set of equations, we 
truncate the hierarchy of  distribution functions via an approximation.
Various methods have been used to break hierarchies resulting from
reaction-diffusion systems \cite{vankampen,lin,kak}.  We shall
restrict ourselves to the simplest. In the hierarchy that begins with
\eqref{rhodot2} and \eqref{gdot2} we make the ansatz
\begin{equation}
   F_3'(y,0^+;t) = g(y,t)g'(0^+,t).
\label{iia}
\end{equation}
This choice, which would be exact if successive interparticle spacings
were independent~\cite{abena}, is not {\em per se} the most compelling,
but it has been shown to produce excellent results (when compared with
simulations) for batch reactions and in the steady state with
unpaired nucleation~\cite{kak,abena}.  In Sec.~\ref{exact} we shall
compare the steady-state density obtained with this closure to the 
exact result.

With the approximation \eqref{iia} we find \(-4D\rho^2(t)F_3'(y,0^+)
=g(y,t)(\dot \rho(t) - 2\Gamma)\), and so \eqref{rhodot2} and \eqref{gdot2}
reduce to the following closed set of equations, linear in \(g(y,t)\):
\begin{eqnarray}
   \dot\rho(t) &=& -4D \rho^2(t)g'(0^+,t) + 2\Gamma,
   \label{rhodott} \\[12pt]
   \frac{\partial}{\partial t}g(y,t) &=&
   2D\frac{\partial^2}{\partial y^2}g(y,t) - 2K(y)g(y,t)+ 
   \frac{4\Gamma}{\rho(t)}(1-g(y,t)) + \frac{q(y)}{\rho^2(t)},
\nonumber \\
   \label{gdott}
\end{eqnarray}
plus the condition \eqref{gcond}.
In the case of paired nucleation, \(q(y) = \Gamma\delta(y-b)\).  In
the case of unpaired nucleation in the (thermodynamic) limit, \(L \gg
\rho(t)^{-1}\), \eqref{gdott} reduces to
\begin{eqnarray}
   \frac{\partial}{\partial t}g(y,t)
   &=& 2D\frac{\partial^2}{\partial y^2}g(y,t) -2K(y)g(y,t)
   + \frac{2Q}{\rho(t)}\left(1-g(y,t)\right),
 \nonumber\\
\label{pairnoaunc}
\end{eqnarray}
with, as before, $Q\equiv2\Gamma$.
Note that no further simplifications can
be obtained by assuming low density. In particular, a low
density expansion cannot be used to justify the truncation \eqref{iia}.

\subsection{Steady states}

The density and correlation function in the steady state, \(\rss\) and
\(\gss(y)\), are found by setting to zero the time derivatives on the
left-hand side of \eqref{rhodott} and of \eqref{gdott} or
\eqref{pairnoaunc}.  There thus
results a second-order equation for \(\gss(y)\), with the two
relations \eqref{gcond} and \eqref{ssdens}.

\begin{figure}
\begin{center}
\leavevmode
\epsfxsize = 4.0in
\epsffile{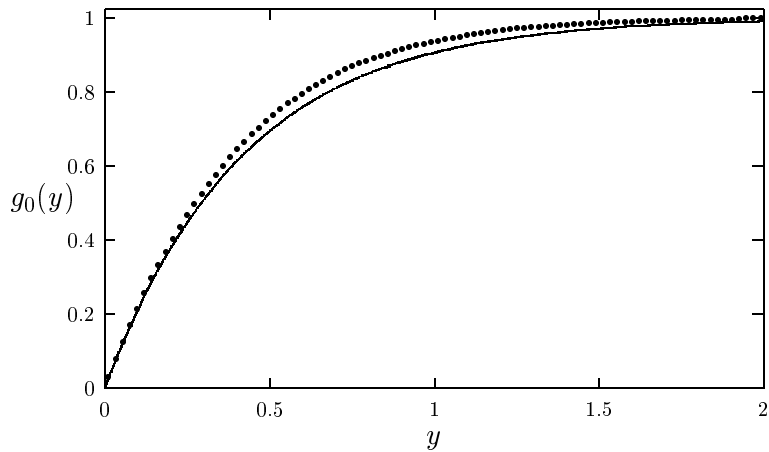}     
\end{center}
\caption{
Correlation function for unpaired nucleation. 
Numerical results are compared with the formula \protect\eqref{ssgunc}
obtained by truncating the hierarchy (solid line).
$Q = 1.0$ and $D = 0.5$.
}
\label{corrun}
\end{figure}

For {\em unpaired nucleation} one finds~\cite{kak}
\begin{eqnarray}
   \rss^{\rm tu} &=& \left(\frac{Q}{16D}\right)^{\frac13},
\label{ssrunc}\\[12pt]
   \gss^{\rm tu}(y) &=& 1 - e^{-2(Q/2D)^{1/3}y},
\label{ssgunc}
\end{eqnarray}
where we have introduced the superscript ``t'' to indicate that the
result is obtained from the truncation \eqref{iia} and ``u'' to denote
``unpaired nucleation."  In Fig.~\ref{corrun} we compare numerical
results for the correlation function with \eqref{ssgunc}. 

\begin{figure}
\begin{center}
\leavevmode
\epsfxsize = 4.0in
\epsffile{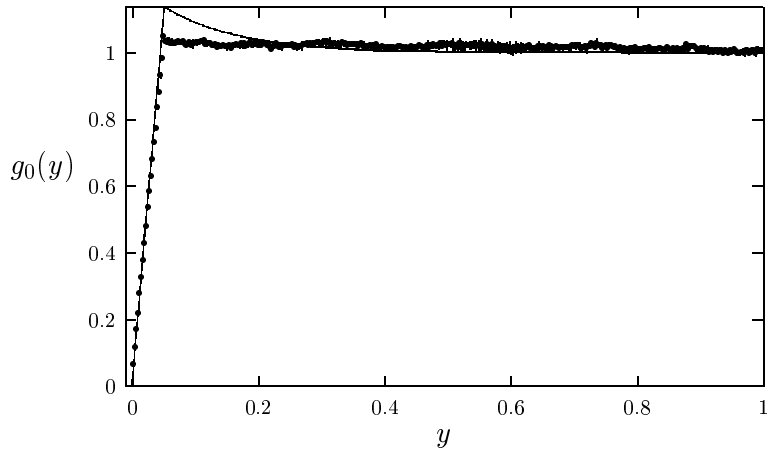}     
\end{center}
\caption{
Correlation function for paired nucleation. 
Numerical results are shown as dots and the
approximation \protect\eqref{corrp} as a solid line.
$b=0.05$, $\Gamma = 64.0$ and $D = 2.0$.
}
\label{corrpfig}
\end{figure}

For {\em paired nucleation} the steady-state equation for the
correlation function is
\begin{eqnarray}
   0 &=&2D\frac{\partial^2}{\partial y^2} \gss(y) -2K(y)\gss(y)
      + 4\frac{\Gamma}{\rss}[1-\gss(y)]
      + \frac{\Gamma}{\rss^2}\delta(y-b).
\label{corr2}
\end{eqnarray}
The solution of \eqref{corr2} is derived in Appendix~A.  When $b$ is
sufficiently large the results are equivalent to \eqref{ssgunc} with
\eqref{ssrunc}.  Of interest here is the opposite situation:
$\eps\rightarrow 0$, with $\epsilon$ defined in Eq.~\eqref{epsdef}. 
The separation \(b\) in the latter case is much
smaller than the length scale defined by the inverse density, and 
the steady-state density is given by
\begin{equation}
   \rss^{\rm tp} = \left(\frac{\Gamma b}{2D}\right)^{1/2}.
\label{corrrho}
\end{equation}
The correlation function in the same limit is
\begin{equation}
   \gss^{\rm tp}(y) = \left\{\begin{array}{lll}
            \frac{\D y}{\D b}
            \qquad &0 \le y < b;\\[15pt]
            1 
            \qquad &y \ge b.            
            \end{array}\right.
\label{corrp}
\end{equation}
Corrections to \eqref{corrp} are proportional
to \(\eps^{\frac34}\).  In Fig.~\ref{corrpfig} this correlation
function is compared with numerical results.

\subsection{Relaxation to the steady state}
\label{relax}

In order to study the relaxation to the steady state, we decompose the
functions $\rho(t)$ and $g(y,t)$ as follows:
\begin{eqnarray}
\label{r1}
\rho(t)&=& \rss + \delta \rho (t) \; , \\ \nonumber\\
g(y,t)&=& \gss(y) + \delta g (y,t) \; ,
\label{g1}
\end{eqnarray}
with $\rss$ and $\gss(y)$ the steady-state density and the
steady-state correlation function, respectively.  This decomposition
is valid for both unpaired
and paired nucleation.  Assuming that we are close to the
steady state, we can obtain linearized equations for the deviations $\delta
\rho$ and $\delta g$ from their steady-state values.  For paired
nucleation
\begin{eqnarray}
\label{lrho1}
\frac{\partial}{\partial t} \delta \rho (t)
&=& - 4 D \rss^2 \delta g'(0^+,t)
- 8 D \rss \gpss(0^+) \delta \rho (t)
\; , \\
\label{lg1}
\nonumber\\
\frac{\partial}{\partial t} \delta g (y,t) &=&  2 D
\frac{\partial^2}{\partial y^2} \delta g (y,t) -2K(y)\delta g(y,t)
\nonumber\\ \nonumber \\
&&-\frac{4 \Gamma}{\rss} \left[ \delta g (y,t) + \frac{1 - \gss (y)}
{\rss} \delta\rho(t) \right]-
\frac{2\Gamma}{\rss^3} \delta(y-b)\delta\rho(t) \; . \nonumber\\
\label{lgho1}
\end{eqnarray}
For unpaired nucleation the last term in \eqref{lgho1} is absent.

Formal solution of these coupled linear equations 
is presented in Appendix~B. Explicit solution for {\em all times} is
in fact possible for the unpaired nucleation case (and presented in the
appendix).  Ultimately we are
interested in their asymptotic relaxation behavior.
If the relaxation processes each involve a single
exponential decay
\begin{eqnarray}
\label{asymptoticrho}
\delta \rho (t) \;& \stackrel{t \rightarrow \infty}{\longrightarrow}&\;
Ae^{- \alpha t}, \\ [12pt]
\delta g'(0^+,t)\;& \stackrel{t \rightarrow \infty}{\longrightarrow}&\;
Be^{- \beta t},
\label{asymptoticg}
\end{eqnarray}
then the density and correlation function decay on the same
time scale, i.e., $\beta = \alpha$. For unpaired nucleation we find
from the exact result (\ref{66}) that the asymptotic decay is indeed
exponential, with
\begin{eqnarray}
   \alpha^{\rm u} &=& (5+\sqrt{5})(D\Gamma^2)^{1/3} = 7.236\ldots
(D\Gamma^2)^{1/3} \nonumber\\ [10pt]
&=& \frac{5+\sqrt{5}}{4^{1/3}} (D Q^2)^{1/3} =
4.558\ldots  (D Q^2)^{1/3}.
\label{unpaireddecay}
\end{eqnarray}

In the case of paired nucleation, if we assume exponential decay
we find for the inverse time scale
\begin{equation}
\alpha^{\rm p} = \left(\frac{32 D\Gamma}{b}\right)^{1/2}.     
\label{paireddecay}
\end{equation}
However, on the basis of the exact results reported below (and also in
parallel work~\cite{avraham}) there is reason to suspect that the decay
may not be purely exponential in the paired nucleation case.

\section{Exact results}
\label{exact}

In this section we derive exact expressions for the density of
particles, using a function that obeys a linear partial differential
equation. The function is similar in interpretation to the pair-pair
correlation function in the Ising model \cite{glauber}. The
methodology is also similar to that used to obtain exact results for
models of diffusion-limited coagulation \cite{da}. Here, we obtain
explicit exact expressions for the density of particles, in steady
state and nonsteady state, for paired and unpaired
nucleation. Previous exact results for diffusion-limited reaction with
annihilation have been limited to the case of no
nucleation~\cite{avraham}.

Let the function \(r(x,t)\) be defined as follows:
\begin{eqnarray}
   r(x,t) = &\{\textrm{probability that the number of particles} 
\label{rdef}
\\&
  \textrm{between } 0 \textrm{ and } x \textrm{ at time }
  t \textrm{ is even}\}.
\nonumber
\end{eqnarray}
Note that, by translational invariance, we can replace
the interval \((0,x)\) by \((X,X+x)\) for any \(X\). 
The value of \(r(x,t)\) changes due to diffusion of particles
in or out of of the region \((0,x)\), and due to nucleation of a single
particle in the region.

In Appendix~C we derive equations for the space and time derivatives of
\(r(x,t)\).  At any time \(t\) the density \(\rho(t)\) is given by
\begin{equation}
   \rho(t) = - \frac{\partial }{\partial x} r(x,t)\big\vert_{x=0^+}.
\label{rhor}
\end{equation}
To describe the time evolution of \(r(x,t)\), we distinguish between the
cases of unpaired and paired nucleation using the superscripts \({\rm
u}\) and \({\rm p}\).

In the case of {\em unpaired nucleation}, \(r^{\rm u}(x,t)\) satisfies

\begin{eqnarray}
   \frac{\partial }{\partial t} r^{\rm u}(x,t) &=&
   2D\frac{\partial^2}{\partial x^2} 
   r^{\rm u}(x,t) - xQr^{\rm u}(x,t) + xQ(1-r^{\rm u}(x,t))\nonumber\\
   \nonumber\\
   &=& 2D\frac{\partial^2}{\partial x^2}
    r^{\rm u}(x,t) + xQ(1 - 2r^{\rm u}(x,t)),
\label{drdt}
\end{eqnarray}
with the boundary conditions
\begin{equation}
   r^{\rm u}(0,t) = 1 \quad \textrm{and}\quad 
   \lim_{x\to\infty}r^{\rm u}(x,t) = \frac12, \qquad t>0.
\label{rbdy}
\end{equation}

In the case of {\em paired nucleation} \(r^{\rm p}(x,t)\) satisfies
\begin{eqnarray}
   \D\frac{\partial }{\partial t} r^{\rm p}(x,t) =
   \left\{\begin{array}{lll}2D\D\frac{\partial^2}{\partial x^2} r^{\rm p}(x,t)
   + 2x\Gamma(1 - 2r^{\rm p}(x,t)) \qquad &x \le b;\\[10pt]
                            2D\D\frac{\partial^2}{\partial x^2} r^{\rm p}(x,t)
   + 2b\Gamma(1 - 2r^{\rm p}(x,t)) \qquad &x > b,
            \end{array}\right.
\label{drdtb}
\end{eqnarray}
with the boundary conditions 
\begin{equation}
   r^{\rm p}(0,t) = 1 \quad \textrm{and}\quad 
   \lim_{x\to\infty}r^{\rm p}(x,t) = \frac12, \qquad t>0.
\label{rbdyp}
\end{equation}

\subsection{Steady state: Unpaired nucleation}

\begin{figure}
\begin{center}
\leavevmode
\epsfxsize = 4.0in
\epsffile{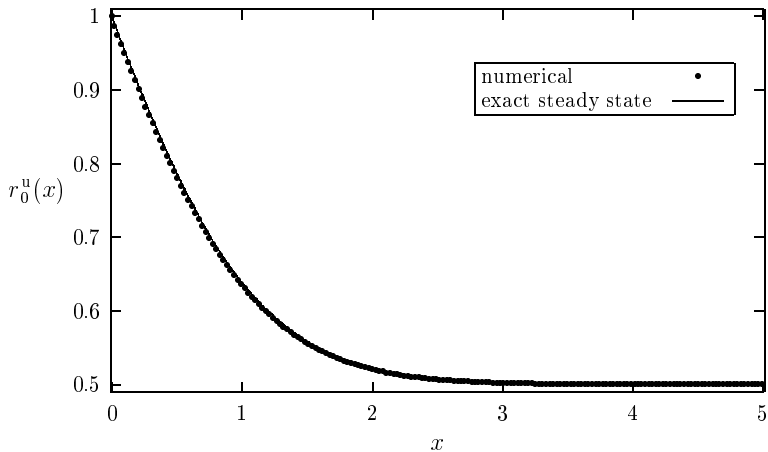}     
\end{center}
\caption{The function \(r_0^{\rm u}(x)\):
numerical and exact results for unpaired nucleation.
The solid line is Eq.~\eqref{rssexact}. \(Q=1.0\) and \(D=0.5\).
}
\label{ru}
\end{figure}

The steady-state solution of \eqref{drdt} will be denoted by 
\(r^{\rm u}_0(x)\). It satisfies
\begin{equation}
   2D\frac{\partial^2}{\partial x^2} r^{\rm u}_0(x) + xQ(1 - 2r^{\rm u}_0(x)) = 0.
\label{ssru}
\end{equation}
The solution is \cite{racz,abena}
\begin{equation}
   r_0^{\rm u}(x) = 
   \frac12\left(1 + \frac{\ai((Q/D)^{\frac13}x)}{\ai(0)}\right)
\label{rssexact}
\end{equation}
and is shown in Fig.~\ref{ru}.
Thus the exact steady-state density for unpaired nucleation is
\begin{equation}
   \rss^{\rm u} = - \frac{\partial }{\partial x}
   r^{\rm u}_0(x)\big\vert_{x=0^+} 
                  = \frac12\left(\frac{Q}{D}\right)^{\frac13}
   \frac{|\ai'(0)|}{\ai(0)} =
\left(\frac{Q}{16D}\right)^{\frac13}0.9186\ldots.  \label{rhossexact}
\end{equation} 
Note that the exact result for the steady-state density
is \(0.9186\) of the density \eqref{ssrunc} predicted from the
truncated hierarchy.  For comparison, in a discrete model with
nucleation rate \(R\), where collision of particles produces
coagulation rather than annihilation, the steady-state density is
given by \cite{da,lin} 
\begin{equation}
   \rss^{\rm u} = 
   \frac12\left(\frac{R}{2D}\right)^{\frac13}\frac{|\ai'(0)|}{\ai(0)}
   \qquad \textrm{(coagulation)}.
\label{sscoag}
\end{equation}

\begin{figure}
\begin{center}
\leavevmode
\epsfxsize = 4.0in
\epsffile{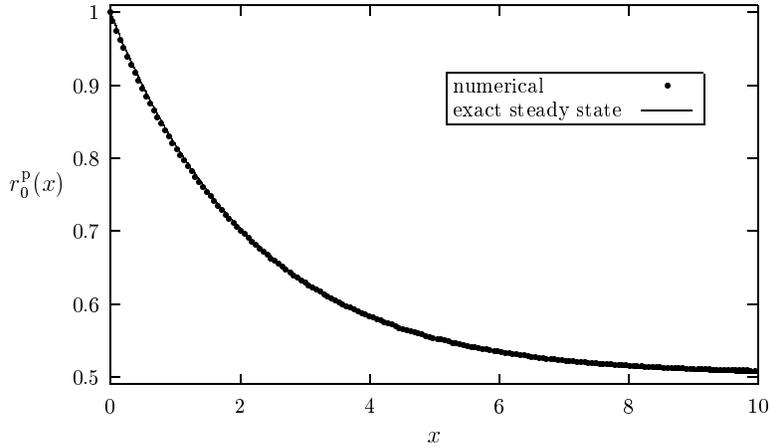}     
\end{center}
\caption{
The function \(r_0^{\rm p}(x)\), measured at late times in a
numerical simulation with paired nucleation (dots). The solid line is
\(r_0^{\rm p}(x)\) as given in Eq.~\eqref{rpsoln}.
\(b=0.2\), \(\Gamma=0.25\) and \(D=0.5\).
}
\label{rp}
\end{figure}

\subsection{Steady state: Paired nucleation}

The steady-state solution of \eqref{drdtb} is
\begin{equation}
   r^{\rm p}_0(x) = \left\{\begin{array}{lll}
     \frac12\left[c_1\ai\left(\alp x\right)
                + c_2\bi\left(\alp x\right) + 1\right]
     \qquad &x \le b;\\[15pt]
     \frac12\left[c_3\exp(-(\frac{2\Gamma b}{D})^{\frac12}x)+1\right]
     \qquad &x > b \end{array}\right.
\label{rpsoln}
\end{equation}
and is shown in Fig.~\ref{rp}.
We have used the second of the boundary conditions \eqref{rbdyp}
to rule out increasing exponential solutions for \(x>b\). The
constants \(c_1\), \(c_2\), and \(c_3\) are fixed by requiring \(r^{\rm
p}_0(0)=0\) and imposing continuity of \(r^{\rm p}_0(x)\) 
and \(\frac{\d}{\d x}\,r^{\rm p}_0(x)\) at \( x=b \).

\begin{figure}
\begin{center}
\leavevmode
\epsfxsize = 4.0in
\epsffile{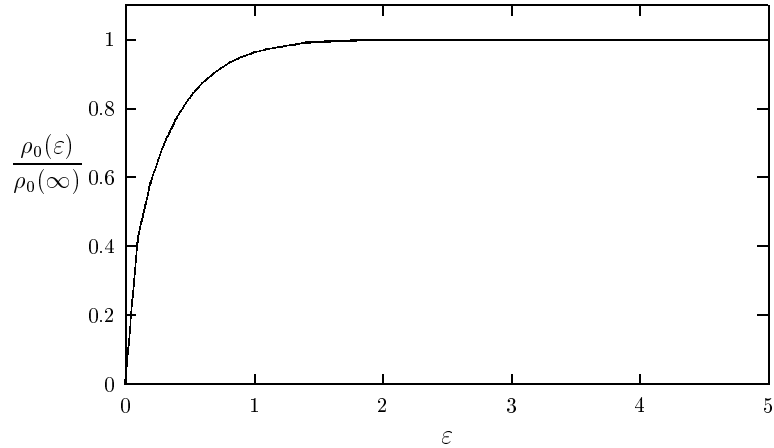}     
\end{center}
\caption{Exact steady-state density versus the
dimensionless parameter \(\eps\).
}
\label{ess}
\end{figure}

The density \(\rho(t)\) is given by \eqref{rhor}. In the
steady state
\begin{eqnarray}
   \rss^{\rm p}
    &=&- \frac{\dd }{\dd x} r_0^{\rm p}(x)\big\vert_{x=0^+}\nonumber\\[12pt]
    &=& \frac12\alp(c_1\ai'(0) + c_2\bi'(0))\nonumber\\[12pt]
    &=&\left(\frac{\Gamma}{4D}\right)^{\frac13}
   \frac{|\ai'(0)|}{\ai(0)}\left(\frac
   {\bi'(\eps) + \surd3\ai'(\eps)
   +\eps(\bi(\eps) + \surd3\ai(\eps))}
   {\bi'(\eps) - \surd3\ai'(\eps)
   +\eps(\bi(\eps) - \surd3\ai(\eps))}
   \right),
\label{rhob}
\end{eqnarray}
in terms of the dimensionless quantity $\eps$ defined in
Eq.~\eqref{epsdef}.
The function \eqref{rhob} is plotted in Fig.~\ref{ess}.

In the limit \(\eps \to 0\), \(\bi(\eps) \to \surd3\ai(\eps)\) and
\(\bi'(\eps) \to -\surd3\ai'(\eps)\), so
\begin{eqnarray}
    \rss^{\rm p} &=& \left(\frac{\Gamma}{4D}\right)^{\frac13}
   \left(\eps^{\frac12} - \frac12\eps^{2} + \ldots\right)\nonumber\\ [12pt]
    &=& \blp(1 + {\cal O}(\eps^{\frac32})).
\label{rhobz}
\end{eqnarray}
For \(\eps\to\infty\)
\begin{equation}
   \rss^{\rm p} \to \frac12\left(\frac{2\Gamma}{D}\right)^{\frac13}
   \frac{|\ai'(0)|}{\ai(0)},
\label{rhobi}
\end{equation}
and we regain the result \eqref{rhossexact}.

\subsection{Time-dependent statistics: Unpaired nucleation}

Let us introduce
\begin{equation}
   h^{\rm u}(x,t) = r^{\rm u}(x,t) - r^{\rm u}_0(x).
\label{hdef}
\end{equation}
Then, \(h^{\rm u}(x,t)\) satisfies
\begin{equation}
   \frac{\partial}{\partial t}h^{\rm u}(x,t) = 
   2D\frac{\partial^2}{\partial x^2} h^{\rm u}(x,t) - 2Qx\,h^{\rm u}(x,t),
\label{hde}
\end{equation}
with the boundary conditions
\begin{equation}
   h^{\rm u}(0,t) = 0 \quad 
\textrm{and}\quad\lim_{x\to\infty} h^{\rm u}(x,t) = 0,
  \qquad
\label{bcu}
\end{equation}
for all $t$.  We can expand the general solution as follows:
\begin{equation}
   h^{\rm u}(x,t) = \sum_{i=1}^{\infty}c_i\,h_i^{\rm u}(x)\ee^{-\lambda_i\,t},
\label{hexp}
\end{equation}
where the eigenfunctions \(h_i^{\rm u}(x)\) satisfy
\begin{equation}
   \frac{\dd^2}{\dd x^2} h_i^{\rm u}(x) - \frac{Q}{D}x\,h_i^{\rm u}(x)
   = -\lambda_i h_i^{\rm u}(x),
\label{hide}
\end{equation}
with the boundary conditions
\begin{equation}
   h_i^{\rm u}(0) = 0 
   \quad \textrm{and} \quad \lim_{x\to\infty} h_i^{\rm u}(x) = 0 .
\label{bcsh}
\end{equation}
The eigenfunctions \(h_i^{\rm u}(x)\) are thus given by
\begin{equation}
   h_i^{\rm u}(x) = N_i\,\ai\left(\left(\frac{Q}{D}\right)^{\frac13}
  (x-\frac{\lambda_i}{2Q})\right).
\label{hisoln}
\end{equation}
The eigenfunctions are normalized by choosing
\begin{equation}
   N_i^{-2} = \int_0^{\infty} \d x \;
\ai^2\left(\left(\frac{Q}{D}\right)^{\frac13}
  (x-\frac{\lambda_i}{2Q})\right) = \left(\frac{D}{Q}\right)^{\frac13}
  \int_{a_i}^{\infty} \d z \; \ai^2(z).
\label{ndef}
\end{equation}
The eigenvalues \(\lambda_i\) are related to the zeros
of the Airy function (all on the negative real axis)
\begin{equation}
   \lambda_i = -(8DQ^2)^{\frac13}a_i,
\label{lambdai}
\end{equation}
where \(a_i\) is the \(i\)th zero counting away from \(0\).
Relaxation towards the steady state is determined for late
times by the smallest eigenvalue
\begin{equation}
   \lambda_1 = -(8DQ^2)^{\frac13}a_1.
\label{lambda1}
\end{equation}

An explicit analytical solution for the density as a function
of time is obtained once the constants \(c_i\) are determined 
from the initial condition \(r^{\rm u}(x,0)\)
\begin{equation}
   c_i = \int_0^{\infty} \d x \;
 h_i^{\rm u}(x)(r^{\rm u}(x,0) - r_0^{\rm u}(x)).
\label{cidef}
\end{equation}
Thus,
\begin{eqnarray}
   h^{\rm u}(x,t) &=&  \sum_{i=1}^{\infty}\frac
   {\int_0^{\infty} \d z \; \ai(z+a_i)
   \left[r(z(\frac{D}{Q})^{\frac13},0) - \frac{\D 1}{\D 2}
   \left(1+\frac{\D \ai(z)}{\D \ai(0)}\right)\right]}
   {{\int_{a_i}^{\infty} \d z \; \ai^2(z)}}
 \nonumber\\[12pt]&& \times 
   \ai\left((\frac{Q}{D})^{\frac13}(x-\frac{\lambda_i}{2Q})\right)
   \ee^{-\lambda_i t},
\label{hsoln}
\end{eqnarray}
and
\begin{eqnarray}
   \rho(t) &=& \rss^{\rm u} 
- \frac{\partial }{\partial x} h^{\rm u}(x,t)\big\vert_{x=0^+}
   \\[12pt]&=& \rss^{\rm u} - \frac12\left(\frac{Q}{D}\right)^{\frac13}
   \sum_{i=1}^{\infty}\frac
   {\int_0^{\infty}  \d z \;
\ai(z+a_i)\left[r^{\rm u}(z(\frac{D}{Q})^{\frac13},0) -
   \frac{\D 1}{\D 2}
   \left(1+\frac{\D \ai(z)}{\D \ai(0)}\right)\right]}
{{\int_{a_i}^{\infty} \d z \;\ai^2(z)}}
   \nonumber\\ [12pt]
   &&\times \ai'(a_i)\ee^{-\lambda_i t}.	
\label{rhot}
\end{eqnarray}

\begin{table}
\begin{center}
\begin{tabular}{|l|| c| c| c| c|} \hline
$i$ & $a_i$ & $\ai'(a_i)$&
$\int_{a_i}^{\infty} \d z \;\ai^2(z)$&
$\int_0^{\infty} \d z \;\ai(z-a_i)(1 - \frac{\ai(z)}{\ai(0)})$
  \\
\hline\hline
&&&&\\
1& -2.338&  0.701& 0.492& 0.972 \\
2& -4.088& -0.803& 0.645& 1.002 \\
3& -5.520&  0.865& 0.749& 0.996 \\
4& -6.786& -0.911& 0.829& 1.001 \\ 
5& -7.944&  0.947& 0.897& 1.012 \\ 
\hline
\end{tabular}
\caption{Quantities related to the eigenvalues
and eigenfunctions for unpaired nucleation.}
\label{tabzic}
\end{center}
\end{table}

\subsubsection{Zero initial density}

If \(\rho(0)=0\), then \(r^{\rm u}(x,0) = 1\) for all $x>0$
and
\begin{equation}
   c_i =    \frac12 N_i \left(\frac{D}{Q}\right)^{\frac13}
   \int_0^{\infty}\d z \; \ai(z+a_i)\left(1 - \frac{\ai(z)}{\ai(0)}\right).
\label{cizic}
\end{equation}
Thus,
\begin{eqnarray}
   \rho(t) &=& \rss^{\rm u} 
- \frac{\partial }{\partial x} h^{\rm u}(x,t)\big\vert_{x=0^+}
   \\[10pt]&=& \rss^{\rm u}
\nonumber
\\
&-&  \frac12\left(\frac{Q}{D}\right)^{\frac13}
   \sum_{i=1}^{\infty}
   \frac{\int_0^{\infty} \d z \; 
\ai(z-a_i)\left(1 - \frac{\D \ai(z)}{\D \ai(0)}   \right)}
   {\int_{a_i}^{\infty}\d z \;\ai^2(z)}\,\ai'(a_i)\ee^{-\lambda_i t}.
\label{rhotzic}
\end{eqnarray}

\begin{figure}
\begin{center}
\leavevmode
\epsfxsize = 4.0in
\epsffile{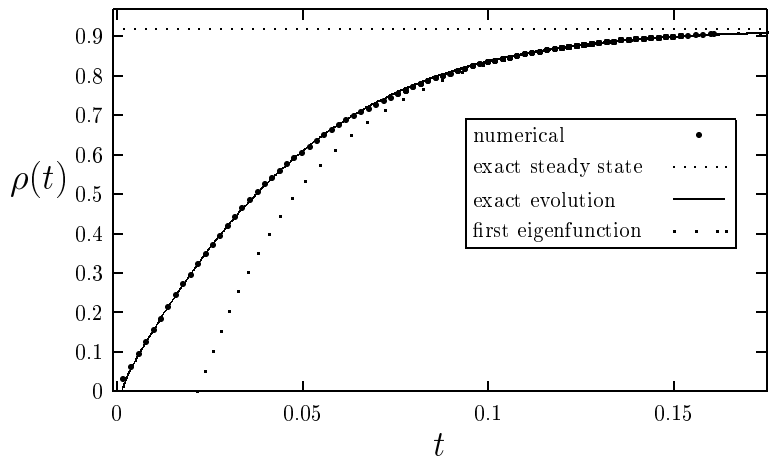}     
\end{center}
\caption{Time evolution starting with no particles present.
Unpaired nucleation with \(Q = 16\), \(D= 1\). 
Solid circles are numerical results.  The solid line, almost
invisible under the numerical results, is the exact evolution calculated
from Eq.~(\ref{rhotzic}).  The upper dotted line is the exact steady
state and the lower dotted line is the first term in the sum
\eqref{rhotzic}.
}
\label{zicfig}
\end{figure}

In Fig.~\ref{zicfig}, the
exact time evolution is compared with numerical
results, obtained with \(L = 3\times10^6\). The lower dotted line is
obtained by plotting only the first term of the sum
in \eqref{rhotzic}, using the values from Table~\ref{tabzic}.
Explicitly, the first eigenvalue that determines the long-time approach
to the steady state is
\begin{equation}
\lambda_1=4.676\ldots (DQ^2)^{1/3} = 7.4227\ldots (D\Gamma^2)^{1/3}.
\end{equation}

\subsubsection{Random initial density} 

An interesting case is provided by starting the system with
the exact steady-state density \(\rho(0)=\rss^{\rm u}\),
but with a random initial distribution of particles.
There is an initial period of rapid annihilation
that reduces the density, followed by a slower relaxation back to the
steady-state value. 

\begin{figure}
\begin{center}
\leavevmode
\epsfxsize = 4.0in
\epsffile{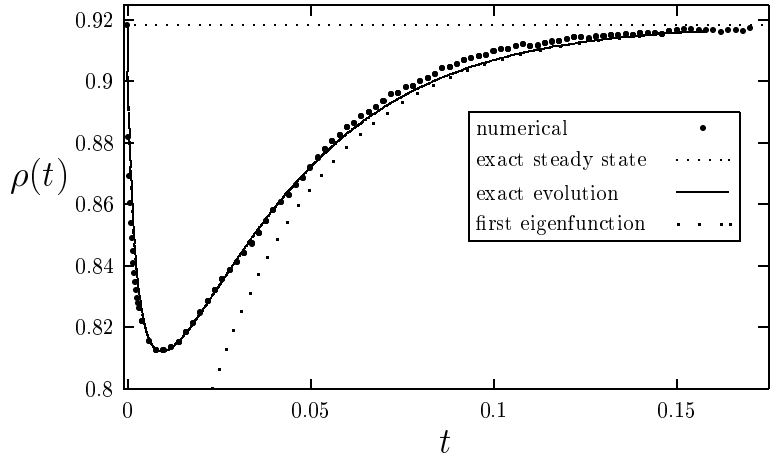}     
\end{center}
\caption{Time evolution starting from a random
distribution of particles with the exact steady state density.
Unpaired nucleation with \(Q = 16\), \(D= 1\).
Solid circles are
numerical results.  The solid line is the exact evolution, 
the upper dotted line is the exact steady state, 
and the lower dotted line is the most slowly decaying
term in the sum \eqref{rhot} with initial conditions
\eqref{rxric}.}
\label{ricfig}
\end{figure}

For a random initial distribution of particles with
density \(\rho\), the number of particles in \((0,x)\) is a
Poisson random variable with mean \(\rho x\). The function 
\(r^{\rm u}(x,0)\) can be calculated as follows:
\begin{eqnarray}
   r^{\rm u}(x,0) &=& 
   \pr{0 \textrm{ particles between } 0 \textrm{ and } x} \nonumber\\
    [12pt]
   &&+ \pr{2 \textrm{ particles between } 0 \textrm{ and } x} + \ldots
   \nonumber\\ [12pt] &=&
   \ee^{-\rho x} + \ee^{-\rho x}\frac{(\rho x)^2}{2! } + \ldots
   \nonumber\\ [12pt] &=& \frac12\left( 1 + \ee^{-2\rho x}\right).
\label{rxric}
\end{eqnarray}
Figure~\ref{ricfig} shows data from a numerical
simulation, performed with \(L=2\times10^5\), along with the results of the
calculation of the coefficients in \eqref{rhot}, using \eqref{rxric}.

\subsection{Time-dependent statistics: Paired nucleation}

Let us introduce
\begin{equation}
   h^{\rm p}(x,t) = r^{\rm p}(x,t) - r^{\rm p}_0(x).
\label{hdefp}
\end{equation}
Then \(h^{\rm p}(x,t)\) satisfies 
\begin{equation}
   \D\frac{\partial }{\partial t} h^{\rm p}(x,t) = {\cal L}\,h^{\rm p}(x,t),
\label{dhdtb}
\end{equation}
where the operator \({\cal L}\) is defined by
\begin{eqnarray}
  {\cal L}\,f(x) =
   \left\{\begin{array}{lll}2D\D\frac{\dd^2}{\dd x^2} f(x)
   - 4x\Gamma f(x) \qquad &x \le b;\\[10pt]
                            2D\D\frac{\dd^2}{\dd x^2} f(x)
   - 4b\Gamma f(x) \qquad &x > b.
            \end{array}\right.
\label{ldef}
\end{eqnarray}
The boundary conditions on \(h^{\rm p}(x,t)\) are
\begin{equation}
   h^{\rm p}(0,t) = 0 \quad \textrm{and} \quad
   h^{\rm p}(x,t) \ \textrm{bounded as}\ {x\to\infty},
\label{bcup}
\end{equation}
for $t>0$.

Let us introduce
\begin{equation}
   \alpha_i = \frac{\lambda_i}{4b\Gamma}.
\label{alphadef}
\end{equation}
For \(\alpha_i > 1\), the eigenvalue equation
\begin{equation}
   {\cal L}h_i^{\rm p}(x) = -\lambda_ih_i^{\rm p}(x),
\label{evep}
\end{equation}
has a continuous spectrum of solutions
\begin{equation}
   h^{\rm p}_i(x) = \left\{\begin{array}{lll}
     c_1(\alpha_i)
\left[\surd3\,\ai\left(\eps (\frac{x}b-\alpha_i)\right) 
                    - \bi\left(\eps (\frac{x}b-\alpha_i)\right)\right]
     \quad &x \le b;\\[15pt]
    c_2(\alpha_i)\sin\left[(\alpha_i-1)^{\frac12}(\frac{2\Gamma b}{D})^{\frac12}(x-b)\right]
    \\[12pt] \qquad
   +~c_3(\alpha_i)\cos\left[ (\alpha_i-1)^{\frac12}(\frac{2\Gamma b}{D})^{\frac12}(x-b)\right]
     \quad &x > b.\end{array}\right.
\label{efpg}
\end{equation}
When \(\eps\) is sufficiently large, there are
also discrete eigenvalues at values of  \(\alpha_i < 1\) satisfying
\begin{eqnarray}
&&   \ai(-\eps\alpha_i)\bi'(\eps(1-\alpha_i))
-\bi(-\eps\alpha_i)\ai'(\eps(1-\alpha_i)) 
   \nonumber \\ [12pt]
&& \quad +~\left(\eps(1-\alpha_i)\right)^{\frac12}\Big(
   \ai(-\eps\alpha_i)\bi(\eps(1-\alpha_i))-\bi(-\eps\alpha_i)\ai(\eps(1-\alpha_i))
   \Big) \nonumber\\ [12pt]
&& \quad =~0.
\label{discalpha}
\end{eqnarray}
The eigenfunctions in this case are
\begin{eqnarray}
   h^{\rm p}_i(x) = \left\{\begin{array}{lll}
     c_1 (\alpha_i)\left[\surd3\,\ai\left(\eps (\frac{x}b-\alpha_i)\right) 
                    - \bi\left(\eps (\frac{x}b-\alpha_i)\right)\right]
     \qquad &x \le b;\\[15pt]
     c_4 (\alpha_i)
\exp[-(\frac{2\Gamma b}{D})^{\frac12}(1-\alpha_i)^{\frac12}x]
     \qquad &x > b.\end{array}\right.
\label{efpl}
\end{eqnarray}

\begin{figure}
\begin{center}
\leavevmode
\epsfxsize = 4.0in
\epsffile{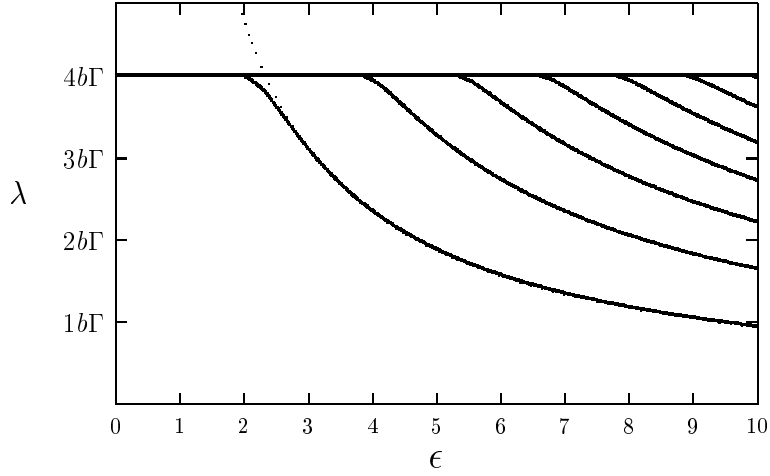}     
\end{center}
\caption{Eigenvalues for paired nucleation.
All values \(\lambda \ge 4b\Gamma\) are permitted.
Discrete values \(\lambda_i < 4b\Gamma\) are also
found for sufficiently large \(\eps\). The dotted line
is \( \lambda=-a_14b\Gamma/\eps\).}
\label{pevlfig}
\end{figure}

Thus, for all finite \(\eps\), there is a
continuous spectrum of eigenvalues with \(\lambda_i \ge 4b\Gamma\). 
For \(\eps\) smaller than \(\eps_c\) these are the only eigenvalues.
The critical value \(\eps_c\) satisfies
\begin{equation}
   \ai(-\eps_c)\bi'(0) -\bi(-\eps_c)\ai'(0) = 0, 
\label{epsceq}
\end{equation}
so
\begin{equation}
   \eps_c = 1.98\ldots.
\label{epscval}
\end{equation}
Discrete eigenvalues appear for larger values of \(\eps\)
(Fig.~\ref{pevlfig}).
The unpaired limit is regained as \(\eps\to\infty\),
when \(-\eps\alpha_i \to a_i\), so that
\(\lambda_i \to  2.338(32D\Gamma^2)^{\frac13}\).

\begin{figure}
\begin{center}
\leavevmode
\epsfxsize = 4.0in
\epsffile{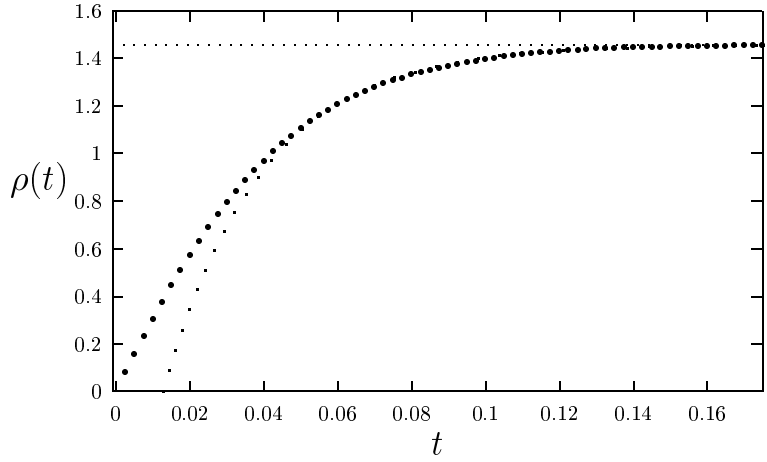}     
\end{center}
\caption{Density of particles versus time for paired
nucleation starting from zero density. The parameters are
\(\Gamma = 16\), \(D= 0.5\) \(b=1\) (\(\eps = 4\)). 
The solid circles are numerical simulation results and
the upper dotted line is the exact steady state. 
The lower dotted curve is \(\rss(1 - 1.6\exp(-2.34b\Gamma t))\). 
Note that the latter exponent is the lowest for \(\eps=4\) and is
a discrete value below the continuous spectrum.
}
\label{eps4fig}
\end{figure}

\begin{figure}
\begin{center}
\leavevmode
\epsfxsize = 4.0in
\epsffile{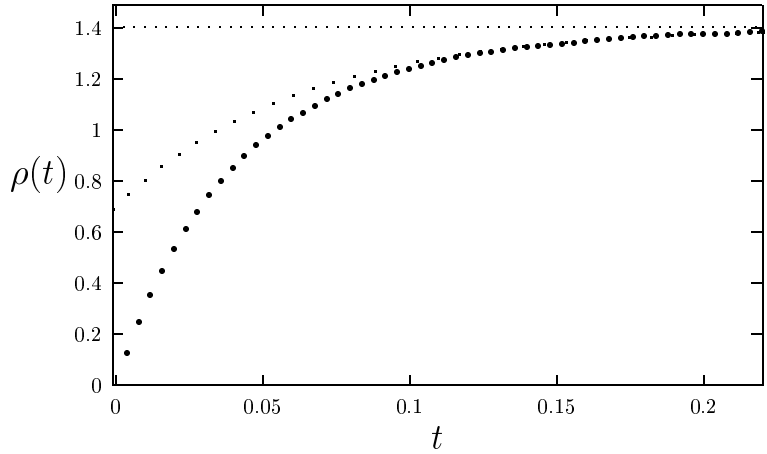}     
\end{center}
\caption{Density of particles versus time for paired
nucleation starting from zero density. The parameters are
\(\Gamma = 16\), \(D= 0.5\) \(b=0.25\) (\(\eps = 1\)).
The solid circles are numerical simulation results, the upper dotted line 
is the exact steady state. 
The lower dotted curve is \(\rss(1 - 0.5\exp(-4b\Gamma t))\). 
The exponent chosen for the fit is the lowest in the continuum.
No discrete eigenvalues are found for \(\eps=1\).}
\label{eps1fig}
\end{figure}

In Figs.~\ref{eps4fig} and \ref{eps1fig} we compare the exact
expressions for the steady-state density and exponents characterizing
relaxation toward the steady-state density with numerical results. In
each case the curved dotted line is obtained using the lowest exponent
available, but the coefficient is obtained from a best fit. In the case
depicted in Fig.~\ref{eps4fig}, there is a discrete eigenvalue
\(\lambda_i < 4b\Gamma\); in the case depicted in
Fig.~\ref{eps1fig}, there is only the continuum of eigenvalues
\(\lambda_i \ge 4b\Gamma\), and the relaxation process may not be
purely exponential.

\subsection{Time-dependent statistics: No nucleation}

In the absence of nucleation, \(r^{\rm n}(x,t)\) satisfies
the heat equation
\begin{eqnarray}
   \frac{\partial }{\partial t} r^{\rm n}(x,t)
    &=& 2D\frac{\partial^2}{\partial x^2}r^{\rm n}(x,t),
\label{drdtnn}
\end{eqnarray}
with the boundary conditions
\begin{equation}
   r^{\rm n}(0,t) = 1 \quad \textrm{and}\quad 
   \lim_{x\to\infty}r^{\rm n}(x,t) = \frac12 ,
\label{rbdynn}
\end{equation}
for all $t>0$.
The solution of \eqref{drdtnn} is given by~\cite{candj}
\begin{eqnarray}
   r^{\rm n}(x,t)  &=& 1 +\left(8\pi Dt\right)^{-\frac12}\Big(
   \int_0^{\infty}\d y\,(1-r^{\rm n}(y,0))\ee^{-(x-y)^2/8Dt}
   \nonumber\\[5pt]&&\qquad\qquad\qquad +
   \int_{-\infty}^0\d y\,(r^{\rm n}(-y,0)-1)\ee^{-(x-y)^2/8Dt}
   \Big).
\label{rsolnnn}
\end{eqnarray}

If the initial distribution of particles is random
with density \(\rho(0)\), then
\begin{equation}
   r^{\rm n}(x,0) = \frac12\left(1 - \ee^{-2\rho(0)x}\right).
\label{rxran}
\end{equation}
Now, using \eqref{rhor}, we derive the density of particles
as a function of time for random initial conditions
\begin{eqnarray}
   \rho(t) &=& - \left(8\pi Dt\right)^{-\frac12}\left(
   \int_0^{\infty}\d
y\,\ee^{-2\rho(0)t}\frac{y}{8Dt}\ee^{-y^2/8Dt}\right.
\nonumber\\ [12pt]
&& \qquad \left.  +
   \int_{-\infty}^0\d y\,\ee^{2\rho(0)t}\frac{y}{8Dt}\ee^{-y^2/8Dt}
   \right)\nonumber\\[12pt]
   &=& 2\rho(0)\left(8\pi Dt\right)^{-\frac12}
   \int_0^{\infty}\d y\,\ee^{-2\rho(0)y}\ee^{-y^2/8Dt}
   \nonumber\\[12pt]
   &=&\rho(0)\exp\left(8Dt\rho^2(0)\right)
   \erfc\left(\rho(0)(8Dt)^{\frac12}\right).
\label{rhodecay}
\end{eqnarray}
We thus reproduce the result \eqref{tandmdens} of
Torney and McConnell~\cite{tandm}.

\section{Discussion}

In the case of { unpaired nucleation}, there is only one
length scale, proportional to \( (D/Q)^{\frac13}\),  
and only one time scale, proportional to \( (DQ^2)^{-\frac13}\).
The relaxation time to equilibrium
and the mean lifetime of a particle are proportional to one another.
This is made clear in the mesoscopic approach~\cite{nucl}, although in
the approaches detailed here we do not make this explicit distinction.
It is noteworthy that the reaction-diffusion approach yields a steady
state that is within 9\% of the correct one and a relaxation rate 
that differs from the exact result by only 2.3\%.

Although the case of paired nucleation is more complicated in terms of
the time scales associated with its dynamics, its steady-state
distribution of particles is closer to a classical equilibrium random
distribution than the corresponding distribution produced by unpaired
nucleation.  Note that the truncated hierarchy approach in this case
leads to the exact steady state density.  The underlying reason is
that the dynamics
produced by paired nucleation is close to time-reversal invariant.
For comparison, an ensemble of noninteracting diffusing particles has
a two-point function \(g(y)\) identically equal to 1 \cite{harris}.
We can imagine producing a spacetime diagram such as shown in
Fig.~\ref{spacetime} from a diagram associated with noninteracting
particles in two steps.  First, when two particles collide, move
them to a different, randomly chosen part of the system. Second,
separate them by a distance \(b\). The first step does not affect the
correlation function or time-reversal invariance.  The second step
directly changes the correlation function for separations smaller than
\(b\). Thus, for diffusion-limited annihilation with paired
nucleation, as the parameter for \(\eps\) that measures the distance
between newly nucleated pairs tends to \(0\), the two-point function
in the steady state is appreciably different from \(1\) only in a
region whose width is proportional to \(b\).

Results obtained for the steady-state
density and relaxation rate are summarized in Table~\ref{tabsummary}.
\begin{table}
\begin{center}
\begin{tabular}{|l|| c| c|| c| c|} \hline
& \multicolumn{2} { c||}{\bfseries Unpaired Nucleation} & \multicolumn{2}
  {c|} {\bfseries Paired Nucleation}\\  \hline
& \multicolumn{2} { c||}{$b\rightarrow \infty $} & \multicolumn{2}
  {c|} {$b\rightarrow 0$}\\  \hline
& $\rho_0$ & Relaxation rate &
  $\rho_0$ & Relaxation rate\\  \hline\hline
Mesoscopic &
$\propto \left(Q/D\right)^{1/3}$ &
$\propto \left(DQ^2\right)^{1/3}$ &
$\propto \left( b \Gamma/D\right)^{1/2}$ & 
$\propto \left( b/\Gamma D\right)^{1/2}$, $ b\Gamma$\\ \hline
Hierarchy &
$\left(Q/16D\right)^{1/3}$ &
$0.219/\left(DQ^2\right)^{1/3}$&
$\left(b\Gamma /2D\right)^{1/2}$ &
$\left(b/32D\Gamma\right)^{1/2}$
 \\ \hline
Exact &
$0.9186\left(Q/16D\right)^{1/3}$ &
$0.2138/\left(DQ^2\right)^{1/3}$ &
$\left(b\Gamma /2D\right)^{1/2}$ &
$ 4 b \Gamma$ \\ \hline
\end{tabular}
\caption{Summary of results for steady state densities and relaxation rates
obtained by various methods.}
\label{tabsummary}
\end{center}
\end{table}
For the evaluation of time scales, the case of { paired nucleation}
is more complicated.  The different approaches indicate the
occurrence of multiple length and time scales.  The mesoscopic
approach~\cite{nucl} leads to a characteristic time for approach to
equilibrium proportional to \( (b\Gamma)^{-1}\), and a distinct mean
lifetime of a particle proportional to \((b/D\Gamma)^{\frac12}\).
These two time scales were identified as corresponding, respectively, to
recombination (two particles created a distance \(b\) apart collide
and annihilate) and to nonrecombinant annihilation (collision between
two particles nucleated at different times). Unpaired annihilation is
less frequent than paired annihilation, but both time scales are
important in the dynamics of the system \cite{nucl}.  The exact
approach leads to the former time scale as an upper bound of a
continuum of scales. The truncated reaction-diffusion hierarchy leads
to the latter time scale under the assumption of exponential decay, which
may not be valid.  Understanding the time scales in the case of paired
nucleation requires further research.

The theoretical approach based on truncation of a hierarchy of
distribution functions thus permits the calculation of steady-state
densities and correlation functions that are in fair agreement with
simulations.  It gives the exact result for the steady-state density
in the limit \(\eps\to 0\), where the statistical distribution of
particles is close to random.  As pointed out by van Kampen
\cite{vankampen}, an approach using a truncation can be made
systematic if it is based on an expansion in a small parameter. However,
his suggestion that the small parameter be the density of particles is
not applicable to the case of unpaired nucleation of point particles
because there is no other quantity with the dimension of length,
i.e., nothing for the density to be small compared to. The exact approach
based on the function \(r(x,t)\) sidesteps these difficulties by
providing a direct method to calculate the density of particles. For
any value of \(\eps\), the density in the steady state and its
time-dependent statistics can be exactly calculated. However, the
method has not yet been extended to exact calculation of the full
distribution of interparticle distances.

\section{Acknowledgments}

The authors acknowledge support from IGPP under Project Los
Alamos/DOE 822AR. K. L. gratefully acknowledges support from the
Engineering Research Program of the Office of Basic Energy Sciences at
the U. S. Department of Energy under Grant
No. DE-FG03-86ER13606. C.M.-P. wishes to acknowledge support from
Centro de Astrobiolog\'{\i}a. S. H., G. L., and C. M.-P. wish to thank
K.L. for her hospitality while visiting USCD, where part of this work
was carried out. The authors are grateful to Ubbo Felderhof
for his insightful comments and corrections, and for the exact inversion of
Eq.~(\ref{11}).

\clearpage

\appendix{\bf APPENDIX A:  Steady state solution of truncated hierarchy}
\label{a}
\setcounter{equation}{0}
\newcounter{appendix}
\setcounter{appendix}{1}
\renewcommand{\theequation}{\Alph{appendix}.\arabic{equation}}

In this Appendix we find the steady state solutions of \eqref{gdott}
and \eqref{pairnoaunc}.

\paragraph{Unpaired nucleation}

We solve \eqref{pairnoaunc} with the left-hand side set to zero.
Fourier transforming the quasilinear equation according to
\begin{equation}
   \hat{g}_n=\frac{2}{L^{1/2}} \int_0^L \d y~\gss(y) 
   \cos \frac{2\pi ny}{L},
\label{ft}
\end{equation}
leads for $n\neq 0$ to
\begin{equation}
   \hat{g}_n=\D{\frac{-\frac{\D 2\Gamma}{\D L^{1/2}\rss^2}}
   {\frac{\D 8D\pi^2 n^2}{\D L^2} +\frac{\D 4\Gamma}{\D \rss}}},
\label{use}
\end{equation}
where Eq.~(\ref{ssdens}) has been used.  Normalization sets
$\hat{g}_0=L^{1/2}$. Fourier inversion according to
\begin{equation}
g(y)=\frac{1}{L^{1/2}}\sum_{n=-\infty}^\infty \hat{g}_n \cos
\frac{2\pi ny}{L},
\end{equation}
can be done
by separating out the $n=0$ contribution explicitly and changing
the resulting sum to an integral (valid as $L\rightarrow\infty$)
\begin{eqnarray}
\gss(y)&=&1-\frac{\Gamma}{\pi \rss^2}\int_0^\infty \d q \
\frac{\D \cos{qy}}{\D q^2+\frac{\D 2\Gamma}{\D D\rss}}\nonumber\\
\nonumber\\
&=& 1-\frac{1}{4}\left(\frac{2\Gamma}{D\rss^3}\right)^{1/2} \ee
^ {-y(2\Gamma/D\rss)^{1/2}}.
\label{ssguncorr}
\end{eqnarray}
The requirement (\ref{gcond}) must also hold in the steady state.
Together with (\ref{ssguncorr}) this leads to
Eqs.~(\ref{ssrunc}) and (\ref{ssgunc}).

\paragraph{Paired nucleation}

We now solve Eq.~\eqref{corr2},  
where  the steady-state density \(\rss\) is related to the
derivative of \(\gss(y)\) as \(y \to 0\)
\begin{equation}
\Gamma = {2D\gpss(0^+)}\rss^2.
\label{ssdensbis}
\end{equation}
The $\delta$-function contribution in the last term of \eqref{corr2}
leads to a discontinuity in the derivative \( \gpss(y)\) at $y=b$.  This
leads to the search for a  solution of the form
\begin{equation}
   \gss(y) = \left\{\begin{array}{lll}
        1 + (S-1)e^{-(2\Gamma/D\rss)^{1/2}y} 
            \qquad &0 \le y < b;\\[15pt]
            1 + Pe^{-(2\Gamma/D\rss)^{1/2}y}
                    + Se^{(2\Gamma/D\rss)^{1/2}y}
            \qquad &y \ge  b.
            \end{array}\right.
\label{corr2soln}
\end{equation}
The constant \(S\) is determined from the condition \eqref{ssdensbis}
\begin{equation}
   1 - 2S = \left(\frac{\Gamma}{8D\rss^3}\right)^{\frac12}.
\label{rsoln}
\end{equation}
Now, the constant \(P\) is determined by enforcing continuity
of the solution \(\gss(y)\) at \(y=b\)
\begin{equation}
   P = S(1 - e^{2(2\Gamma/D\rss)^{1/2} b}) - 1.
\label{psoln}
\end{equation}
The discontinuity in the derivative \(\gpss(y)\) at $y=b$ is
\begin{equation}
   \gpss(b^+)-\gpss(b^-) =  -\frac{\Gamma}{2D\rss^2}.
\label{dgp}
\end{equation}
Using \eqref{corr2soln} to evaluate the left-hand side of
\eqref{dgp} and rearranging gives an implicit expression for the
steady-state density
\begin{equation}
   \left(\frac{\Gamma}{8D}\right)^{1/2} =    \left(
\left(\frac{\Gamma}{8D}\right)^{1/2}
   - \rss^{3/2}\right)e^{(2\Gamma/D\rss)^{1/2}b},
\label{dgpre1}
\end{equation}
or, rearranging again
\begin{equation}
   (b\rss)^{3/2} =
   \frac{\sigma^2}{8}\left(1-\ee^{-\sigma^2/2(b\rss)^{1/2}}\right),
\label{dgpre}
\end{equation}
where \(\sigma\) is defined by
\begin{equation}
   \sigma = \left(\frac{8b^3\Gamma}{D}\right)^{\frac14} \equiv
\sqrt{2}\eps^{3/4}.
\label{sigma}
\end{equation}

While we cannot invert \eqref{dgpre} explicitly for the steady-state
density, we can examine the limits in the dimensionless parameter
\(\sigma\).  In the limit $\sigma\gg 1$, corresponding
to large initial separation, we find
\begin{equation}
   (b\rss)^{3/2} \to \frac{\sigma^2}{8},
\label{bigeps}
\end{equation}
so \(\rss^3 \to \Gamma/8D\), as obtained in \eqref{ssrunc} for
the unpaired nucleation case.  Morevover, \(S \to 0\) and \(g(y) \to
1 - \ee^{-4\rss y}\) as in \eqref{ssgunc}.

The limit $\sigma \rightarrow 0$ corresponds to small
initial separation. Then \eqref{dgpre} reduces to
\begin{equation}
   b\rss = \frac{\sigma^2}{4}
   \left(1 - \frac14\sigma + {\cal O}(\sigma^2)\right).
\label{corrnondim}
\end{equation}
Note that \(\sigma\to 0\) corresponds to \(b\rss \to 0\).
Expanding in powers of \(\sigma\), we find
\begin{equation}
   \rss = \left(\frac{b\Gamma}{2D}\right)^{\frac12}
   (1 - \frac14\sigma + {\cal O}(\sigma^2)),\quad
   S = -\frac12\,\sigma^{-1} + \frac5{16} + {\cal O}(\sigma)
   \quad\textrm{and}\quad
   P = \frac12\,\sigma + {\cal O}(\sigma^2).
\label{smallsig}
\end{equation}
The correlation function \(\gss\) can be expanded as
\begin{equation}
   \gss(y) = \left\{\begin{array}{lll}
            \frac{\D y}{\D b}(1+\frac12\,\sigma) + {\cal O}(\sigma^2)
            \qquad &0 \le y < b;\\[15pt]
            1 + \frac1{2}\sigma e^{-\sigma y/b} + {\cal O}(\sigma^2)
            \qquad &y \ge b.
            \end{array}\right.
\label{corrpsig}
\end{equation}
These results as $\sigma\rightarrow 0$ are reported in
Eqs.~\eqref{corrrho} and \eqref{corrp}.

\clearpage

\appendix{\bf APPENDIX B: Relaxation to the Steady State}
\label{b}

\setcounter{equation}{0}
\setcounter{appendix}{2}
\renewcommand{\theequation}{\Alph{appendix}.\arabic{equation}}

Here we detail the calculation of approach to the steady state in the
truncated hierarchy approach.

\paragraph{Unpaired nucleation} 

It is convenient to introduce the symbols
\begin{equation}
\chi = (D\Gamma^2)^{1/3}, \qquad \gamma= (D^2 \Gamma)^{1/3}.
\label{symbols}
\end{equation}
The linearized perturbation equations \eqref{lrho1} and \eqref{lgho1}
in the unpaired case are
\begin{eqnarray}
\label{lrho1u}
\frac{\partial}{\partial t} \delta \rho (t) &=&
- \chi \left[ \delta g'(0^+,t) +8\delta\rho(t)\right], \\
\label{lg1u}
   \nonumber\\
   \frac{\partial}{\partial t} \delta g (y,t) &=&  2 D
   \frac{\partial^2}{\partial y^2} \delta g (y,t) 
    -2K(y)\delta g(y,t)
   -8\chi\delta g(y,t) \nonumber\\ [10pt]
   && -16\gamma\left[1-\gss^{\rm
   u}(y)\right] \delta\rho(t),
\label{dgpdeu}
\end{eqnarray}
where the specific form \eqref{ssrunc} has been implemented.
As initial conditions we choose
\begin{equation}
\delta g(y,t=0) = 0,
\end{equation}
and an arbitrary $\delta \rho(0)$.  Note that
having implemented the condition \eqref{glim}
on the steady-state solution implies that $\delta \hat g_{n=0} (t) = 0$.

The solution of Eq.~(\ref{lrho1u}) is 
\begin{equation}
\label{lrho2u}
\delta \rho (t) = e^{-8\chi t}\delta \rho(0) -\chi \int_0^t \d \tau
e^{-8\chi(t-\tau)} \delta g'(0^+,\tau).
\end{equation}
Transforming \eqref{dgpdeu} according to \eqref{ft} gives, for
$n \neq 0$
\begin{equation}
\label{fg1}
\frac{\partial}{\partial t} \delta \hat g_n (t) =
- \frac{8 D \pi^2 n^2}{L^2} \delta \hat g_n (t)
-8\chi \delta \hat g_n (t)
- \frac{8 D}{L^{1/2}} \delta g'(0^+,t)
+ 16 \gamma\hat g_n \delta \rho (t)  .
\end{equation}
We can formally solve this equation as well, to obtain
\begin{equation}
\delta \hat g_n (t) = \int_0^t \d \tau
e^{-8 \left(\frac{ D \pi^2 n^2}{L^2} +
\chi \right)(t-\tau)}
\left[
- \frac{8 D}{L^{1/2}}  \delta g'(0^+,\tau) + 16\gamma \hat g_n
\delta \rho (\tau) \right],
\label{lg2}
\end{equation}
and hence its inverse Fourier transform
\begin{equation}
\label{lg5}
\delta  g (y,t) =
-  \int_0^t \d \tau \;  {\cal K} (y, t- \tau) \delta g'(0^+,\tau)
+ 8 \int_0^t \d \tau  \; {\cal G}(y,t-\tau) \delta \rho (\tau).
\end{equation}
We have introduced the following functions
(and taken the limit $L\rightarrow \infty$):
\begin{eqnarray}
{\cal K} (y,t)&=& \frac{8D}{L} e^{-8\chi t}
{\sum'}_{n=-\infty}^{+ \infty }
\cos{\frac{2 \pi  n y}{L}}\;
e^{- \frac{8 D \pi^2 n^2}{L^2} t} \nonumber\\ [12pt]
&=& \frac{8D}{(2\pi Dt)^{1/2}} e^{-8\chi t} e^{-\frac{y^2}{2Dt}},
\label{deltaK}
\\ [12pt]
{\cal G}(y,t)&=&
 \frac{2\gamma}{L^{1/2}} e^{-8\chi t}
{\sum'}_{n=-\infty}^{+ \infty }
\cos{\frac{2 \pi n y}{L}} \;
e^{{-} \frac{8 D \pi^2 n^2}{L^2} t} \; \hat g_n
\nonumber\\ [12pt]
&=& -\frac{8\gamma^2}{\pi}
e^{-8\chi t}
\int_{-\infty}^\infty \; {\rm d}q \;
e^{-2Dq^2t}
\frac {1}{8\chi +2Dq^2} \; \cos qy ,
\label{deltaG}
\end{eqnarray}
where the prime on the sums indicates omission of the $n=0$ term and
we have used Eq.~(\ref{use}).

Since the unknowns appear on both sides of \eqref{lrho2u} and
\eqref{lg5}, these are only formal solutions.  To proceed, we Laplace
transform them (indicated by a tilde, ${\cal L} f(t) = \tilde{f}(s)$,
and solve the resulting set
self-consistently.  The limit \eqref{gzddef} must be
handled carefully and not implemented prematurely.  We find
\begin{eqnarray}
\label{first}
\delta \tilde{\rho} (s) &=&
\frac {1}{8\chi (u+1)} \left(\delta \rho(0) -\chi
\delta\tilde{g}'(0^+,s)\right),
\\ [12pt]
\delta \tilde{g}(y,s) &=& - \tilde{{\cal K}}(y,s)
\delta\tilde{g}'(0^+,s)
+8\tilde{{\cal G}}(y,s) \delta \tilde{\rho}(s).
\label{second}
\end{eqnarray}
From these two equations we can obtain an expression for $\delta
\tilde{g}'(0^+,s)$ as follows.  First, set $y=0$ in \eqref{second}.
Since $g(0,t)=0$ and since $g_0(0)=0$, it follows from Eq.~(\ref{g1}) that 
$\delta g(0,t)=0$ for all $t$ and therefore $\delta \tilde{g}(0,s)=0$ for
all $s$.  Thus, we find
\begin{equation}
0 = - \tilde{{\cal K}}(0,s)
\delta\tilde{g}'(0^+,s)
+8\tilde{{\cal G}}(0,s) \delta \tilde{\rho}(s).       
\end{equation}
Subsitution of \eqref{first} into this result immediately leads to
\begin{equation}
\chi \delta \tilde{g}'(0^+,s)=
\frac{ \D \delta \rho (0)\tilde{\cal G}(0,s)} {\D (u+1) 
\tilde{\cal K}(0,s) + \tilde{\cal
G}(0,s)},
\label{haveitall}
\end{equation}
where we have set $u\equiv s/8\chi$.
This, together with Eqs.~(\ref{first}) and (\ref{second}),
constitutes a complete solution of the Laplace transform of the
problem.  The indicated transforms can be calculated explicitly
\begin{eqnarray}
\tilde{\cal K}(0,s) &=& \frac{ 2(D/\Gamma)^{1/3}}{\sqrt{u+1}},
\\ [12pt]
\tilde{\cal G}(0,s) &=& -\frac{(D/\Gamma)^{1/3}}{4u} \left(
1-\frac{1}{\sqrt{u+1}}\right),
\end{eqnarray}
and one readily obtains
\begin{equation}
\delta \tilde{\rho}(s) = \frac{\delta \rho(0)}{\chi}
\left(\frac{u}{1-\sqrt{1+u}+8u(1+u)}\right).
\label{11}
\end{equation}
It is possible to Laplace invert these expressions
exactly~\cite{private}. In particular, Eq.~(\ref{11}) can be rewritten as
\begin{equation}
\delta \tilde{\rho}(s) = \frac{\delta \rho(0)}{\chi} \left( 
\frac{A_1}{\sqrt{u+1}-y_1} +
\frac{A_2}{\sqrt{u+1}-y_2} +
\frac{A_3}{\sqrt{u+1}-y_3} \right).
\label{22}
\end{equation}
where
\begin{equation}
A_1= \frac{5+\sqrt{5}}{40} \qquad
A_2= \frac{5-\sqrt{5}}{40} \qquad
A_3 = -\frac{1}{4}
\label{33}
\end{equation}
and
\begin{equation}
y_1 = \frac{1}{4}\left( \sqrt{5} -1\right) \qquad
y_2 = - \frac{1}{4}\left( \sqrt{5} +1\right) \qquad
y_3 = -\frac{1}{2}.
\label{44}
\end{equation}
The inversion
\begin{equation}
{\cal L}^{-1} \left( \frac{1} {\sqrt{s+A} -B}\right) = e^{-At} \left(
\frac{1}{\sqrt{\pi t}} + Be^{B^2t} {\rm erfc} (-B\sqrt{t})\right)
\label{55}
\end{equation}
can then be applied to obtain
\begin{equation}
\delta\rho(t) =  8 \delta\rho(0) e^{-8\chi t} \sum_i A_i y_i e^{8\chi y_i^2
t} {\rm erfc} (-y_i\sqrt{8\chi t} ).
\label{66} 
\end{equation}
It is noteworthy that this solution is {\em exact}
within the truncation approximations of the model, that is,
it represents  the full time-dependent solution for the
model.

Asymptotic analysis of the exact result yields pure exponential decay
as indicated in Eqs.~(\ref{asymptoticrho}) and (\ref{asymptoticg}),
with
\begin{equation}
\alpha=(5+\sqrt{5}) \chi = 7.236\dots \chi .
\end{equation}
The proportionality of $\delta \tilde{g}$ and $\delta\tilde{\rho}$
clearly leads to the same decay rate for $\delta g(y,t)$ as for
$\delta\rho(t)$.

\paragraph{Paired nucleation}

Here it is convenient to introduce the symbols
\begin{equation}
\Omega = \frac{4\Gamma}{\rho_0}, \qquad \theta=\Gamma b.
\label{symbols--paired}
\end{equation}
The linearized perturbation equations \eqref{lrho1} and \eqref{lgho1}
in the paired case are
\begin{eqnarray}
\label{lrho1u--paired}
\frac{\partial}{\partial t} \delta \rho (t) &=&
- 4 D \rho_0^2 \delta g'(0^+,t) - \Omega \delta\rho(t), \\
\label{lg1u--paired}
\nonumber\\
\frac{\partial}{\partial t} \delta g (y,t) &=&  2 D
\frac{\partial^2}{\partial y^2} \delta g (y,t) 
 - 2K(y) \delta g(y,t) - \Omega \delta g(y,t)\nonumber\\ [12pt]
&&-\frac{4 \Gamma }{\rho_0}\left[1-\gss^{\rm
p}(y)\right] \delta\rho(t) - \frac{2 \Gamma}{\rho_0^3} \delta (y-b)
\delta \rho(t).
\label{xxx--paired}
\end{eqnarray}

With the same initial condition as in the unpaired case,
the solution of Eq.~(\ref{lrho1u--paired}) is formally given by
\begin{equation}
\label{lrho2u--paired}
\delta \rho (t) = e^{-\Omega t}\delta \rho(0) -4 D \rho_0^2 \int_0^t \d \tau
e^{-\Omega(t-\tau)} \delta g'(0^+,\tau).
\end{equation}
Transforming \eqref{xxx--paired} according to \eqref{ft} gives for
$n \neq 0$
\begin{eqnarray}
\frac{\partial}{\partial t} \delta \hat g_n (t) &=&
- \frac{8 D \pi^2 n^2}{L^2} \delta \hat g_n (t)
-\Omega \delta \hat g_n (t)
- \frac{8 D}{L^{1/2}} \delta g'(0^+,t)\nonumber\\ [12pt]
&& +  \frac{\Omega}{\rho_0}\hat g_n \delta \rho (t) 
-  \frac{4 \Gamma}{\rho_0^3 L^{1/2}} \cos \left( \frac{2 \pi n b}{L}\right) 
\delta \rho (t) .
\label{fg1--paired}
\end{eqnarray}
We can formally solve this equation as well, to obtain 
\begin{eqnarray}
\delta \hat g_n (t) &=& \int_0^t \d \tau 
 e^{-\left(\frac{ 8D \pi^2 n^2}{L^2} +
\Omega \right)(t-\tau)}
\nonumber\\
&&  \times
\left\{
- \frac{8 D}{L^{1/2}}  \delta g'(0^+,\tau) +  
\left[
\frac{\Omega}{\rho_0} \hat g_n - 
 \frac{4 \Gamma}{\rho_0^3 L^{1/2}} \cos \left( \frac{2 \pi n b}{L}\right) 
\right]
\delta \rho (\tau) \right\},
\nonumber
\end{eqnarray}
and hence its inverse Fourier transform (notice that $\delta g_{n=0}(t)=0$
for all $t$)
\begin{eqnarray}
\delta  g (y,t) &=&
-  \int_0^t \d \tau \;  {\cal K} (y, t- \tau) \delta g'(0^+,\tau)
 + \int_0^t \d \tau  \; {\cal G}(y,t-\tau) \delta \rho (\tau)
   \nonumber\\ [12pt]
&&
-\frac{4 \Gamma}{\rho_0^3 L^{1/2}} 
 \int_0^t \d \tau  \; {\cal H}(y,t-\tau) \delta \rho (\tau) .
\label{lg5--paired}
\end{eqnarray}
We have introduced the following functions:
\begin{eqnarray}
{\cal K} (y,t)&=& \frac{8D}{L} e^{-\Omega t} {\sum'}_{n=-\infty}^{+
\infty }
\cos{\frac{2 \pi  n y}{L}}\;
e^{- \frac{8 D \pi^2 n^2}{L^2} t} \nonumber\\ [12pt]
&=& \frac{8D}{(2\pi Dt)^{1/2}} e^{-\Omega t} e^{-\frac{y^2}{2Dt}},
\label{deltaK--paired}
\\ [12pt]
{\cal G}(y,t)&=&
 \frac{\Omega}{\rho_0L^{1/2}} e^{-\Omega t} {\sum'}_{n=-\infty}^{+ \infty }
\cos{\frac{2 \pi n y}{L}} \;
e^{{-} \frac{8 D \pi^2 n^2}{L^2} t} \; \hat g_n,
\label{deltaG-paired}
\\ [12pt]
{\cal H}(y,t)&=&
 \frac{1}{L^{1/2}} e^{-\Omega t} {\sum'}_{n=-\infty}^{+ \infty }
\cos{\frac{2 \pi n y}{L}} \; \cos{\frac{2 \pi n b}{L}} \;
e^{{-} \frac{8 D \pi^2 n^2}{L^2} t} .
\label{deltaH-paired}
\end{eqnarray}

We Laplace transform \eqref{lrho2u--paired} and \eqref{lg5--paired}
and solve the resulting set
self-consistently, to obtain
\begin{eqnarray}
\delta \tilde{\rho} (s) &=&
\frac {1}{ s+\Omega} \delta \rho(0) -4 D \rho_0^2
\frac{\delta\tilde{g}'(0^+,s)}{s+\Omega},
\nonumber
\\ [12pt]
\delta \tilde{g}(y,s) &=& 
- \tilde{{\cal K}}(y,s) \delta\tilde{g}'(0^+,s)
+\tilde{{\cal G}}(y,s) \delta \tilde{\rho}(s)
-\frac{4 \Gamma}{\rho_0^3 L^{1/2}} 
\tilde{{\cal H}}(y,s) \delta \tilde{\rho}(s),
\nonumber
\end{eqnarray}
from which in turn we find
\begin{eqnarray}
\delta \tilde{g}(y,s) &=&
- \tilde{{\cal K}}(y,s) \delta\tilde{g}'(0^+,s)
+ \left[ 
\tilde{{\cal G}}(y,s) 
-\frac{4 \Gamma}{\rho_0^3 L^{1/2}} 
\tilde{{\cal H}}(y,s) 
\right]
\nonumber\\ [12pt]
&& \times
 \left[ 
\frac {1}{ s+\Omega} \delta \rho(0) -4 D \rho_0^2
\frac{\delta\tilde{g}'(0^+,s)}{s+\Omega}
\right].
\label{haveitall--paired}
\end{eqnarray}

We are interested in obtaining $\delta\tilde{g}'(0^+,s)$.  This can 
readily be done by evaluating the previous equation at $y=0$, to
obtain (notice that we have chosen $\delta g(y=0,t)=0$ for all $t$)
\begin{eqnarray}
0
 &=& 
- \tilde{{\cal K}}(0,s) \delta\tilde{g}'(0^+,s)
+ \left[ 
\tilde{{\cal G}}(0,s) 
-\frac{4 \Gamma}{\rho_0^3 L^{1/2}} 
\tilde{{\cal H}}(0,s) 
\right]
\nonumber\\ [12pt]
&& \times 
 \left[ 
\frac {1}{ s+\Omega} \delta \rho(0) -4 D \rho_0^2
\frac{\delta\tilde{g}'(0^+,s)}{s+\Omega}
\right].
\label{haveitall--paired--atzero}
\end{eqnarray}

The previously introduced functions evaluated at the origin become
\begin{eqnarray}
{\cal K} (0,t)&=&
 \frac{8D}{(2\pi Dt)^{1/2}} e^{-\Omega t} ,
\label{deltaK--paired--atzero}
\\ [12pt]
{\cal G}(0,t)&=&
 \frac{\Omega}{\rho_0L^{1/2}} e^{-\Omega t} {\sum'}_{n=-\infty }^{+ \infty }
e^{{-} \frac{8 D \pi^2 n^2}{L^2} t} \; \hat g_n,
\label{deltaG-paired-atzero}
\\ [12pt]
{\cal H}(0,t)&=&
 \frac{1}{L^{1/2}} e^{-\Omega t} {\sum'}_{n=-\infty 
}^{+ \infty }
\cos{\frac{2 \pi n b}{L}} \;
e^{{-} \frac{8 D \pi^2 n^2}{L^2} t} .
\label{deltaH-paired-atzero}
\end{eqnarray}
We are interested in obtaining the Laplace transform of these
functions. It is easy to obtain
\begin{eqnarray}
\tilde{\cal K}(0,s) &=& \frac{ 2(D/\Gamma)^{1/3}}{\sqrt{s/\Omega+1}},
\\ [12pt]
{\cal H}(0,t) &=&
\frac{L^{1/2}}{\pi} e^{-\Omega t} \int_0^{+\infty} \d q \;
e^{-2 D q^2t} \; \cos qb,
\end{eqnarray}
so that
\begin{eqnarray}
\frac{4 \Gamma}{\rho_0^3 L^{1/2}}{\cal H}(0,t) &=&
\frac{2 \Gamma}{\rho_0^3 \sqrt{2 \pi D t}} e^{-\Omega t} e^{-b^2/(8Dt)},
\\ [12pt]
\frac{4 \Gamma}{\rho_0^3 L^{1/2}}\tilde{\cal H}(0,s) &=&
\frac{2 \Gamma}{\rho_0^3 \sqrt{2 D}} 
\frac{1}{\sqrt{s+\Omega}}
e^{-b \sqrt{\frac{s+\Omega}{2D}}}.
\end{eqnarray}

The Fourier components for the two-point correlation function 
associated with Eq.~(\ref{corr2soln}) are
\begin{eqnarray}
\hat g_n&=& \frac{2}{L^{1/2}} \frac{\sqrt{2\Gamma/(D\rho_0)}}{\frac{2 \Gamma}
{D \rho_0}+\frac{4 \pi^2 n^2}{L^2}}
\left[
(2 S -1) - 2S e^{b \sqrt{2\Gamma/(D\rho_0)}} \cos \frac{2 \pi n b}{L}
\right].
\end{eqnarray}
This form in turn leads to the Laplace transform of ${\cal G}(0,t)$: 
\begin{eqnarray}
\tilde{\cal G} (0,s)&=&
 \frac{\Omega}{\pi} \sqrt{\frac{2\Gamma}{D\rho_0^3}}
2 (2S-1) \int_0^{+\infty} \d q
\frac{1}{(q^2 + 2\Gamma/(D \rho_0))(s+\Omega + 2Dq^2)}
\nonumber
\\ [12pt]
&& - \frac{\Omega}{\pi} \sqrt{\frac{2\Gamma}{D\rho_0^3}}
2S e^{b\sqrt{2\Gamma/(D\rho_0)}}
\nonumber\\ [12pt]
&& \times
2 \int_0^{+\infty} \d q
\frac{\cos qb}{(q^2 + 2\Gamma/(D \rho_0))(s+\Omega + 2Dq^2)}.
\end{eqnarray}

It is now a straightforward matter to 1) collect the various Laplace
transform expressions to solve for $\delta\tilde{g}'(0^+,s)$ using
Eq.~(\ref{haveitall--paired--atzero}), 2) substitute this result into 
Eq.~(\ref{first}), and explore the poles of the denominator of the
resulting $\delta \tilde{\rho}(s)$ in the limit $\epsilon \rightarrow
0$.  The procedure is tedious but leads to the inverse time scale
\eqref{paireddecay}.  
The proportionality of $\delta \tilde{g}$ and $\delta\tilde{\rho}$
clearly leads to the same decay rate for $\delta g(y,t)$ as for
$\delta\rho(t)$.  We do note that it is not clear from this procedure
that the relaxation process is actually exponential in time.  If it {\em
is} exponential (and there is reason to question this from the results
of the exact and mesoscopic procedures), then it is necessary to
perform the inverse Laplace transform more carefully.  This is possible,
but beyond the scope of this paper.

\clearpage

\appendix{\bf APPENDIX C: The function} \(r(x,t)\) {\bf and its
derivatives}
\label{c}
 
\setcounter{equation}{0}
\setcounter{appendix}{3}
\renewcommand{\theequation}{\Alph{appendix}.\arabic{equation}}

\begin{figure}
\begin{center}
\leavevmode
\epsfxsize=6.0in
\epsffile{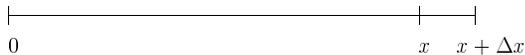}
\vspace{-6.5in}
\end{center}
\caption{The intervals $(0,x)$ and $(x,x+\delx)$.}
\label{intervals}
\end{figure}

We first calculate the derivative of \(r(x,t)\) with respect to \(x\) by
considering the intervals shown in Fig.~\ref{intervals}.
 
\begin{itemize}
\item
Let ${\mathcal P}_{{\rm e}0}(x,\delx,t)$ be the probability that there is
an even number
of particles in \((0,x)\) and no particle in \((x,x+\delx)\) at time
\(t\).
\item
Let ${\mathcal P}_{{\rm e}1}(x,\delx,t)$ be the probability that there is
an even number
of particles in \((0,x)\) and one particle in \((x,x+\delx)\) at time
\(t\).
\item
Let ${\mathcal P}_{{\rm o}1}(x,\delx,t)$ be the probability that there is
an odd number of particles in \((0,x)\) and one particle in
\((x,x+\delx)\) at time \(t\).
\end{itemize}
 
The function \(r(x,t)\) defined in Eq.~(\ref{rdef}) can be expressed in
terms of these quantities as follows:
\begin{eqnarray}
   r(x+\delx,t) &=&{\mathcal P}_{{\rm e}0}(x,\delx,t)
   + {\mathcal P}_{{\rm o}1}(x,\delx,t) + \Or{\delx^2}\nonumber \\ [12pt]
   r(x,t) &=& {\mathcal P}_{{\rm e}0}(x,\delx,t)
   + {\mathcal P}_{{\rm e}1}(x,\delx,t) + \Or{\delx^2}.
\label{rxrxdx}
\end{eqnarray}
Thus
\begin{equation}
   r(x+\delx,t) - r(x,t) =
    {\mathcal P}_{{\rm o}1}(x,\delx,t) - {\mathcal P}_{{\rm e}1}(x,\delx,t)
    + \Or{\delx^2}.
\label{drx}
\end{equation}
In particular, by choosing \(x=0\),
\begin{eqnarray}
   r(\delx,t) - r(0,t) &=&
   0 - {\mathcal P}_{{\rm e}1}(0,\delx,t) + \Or{\delx^2}\nonumber\\ [12pt]
   &=& -\Delta x \rho(t) + \Or{\delx^2}.
\label{rzt}
\end{eqnarray}
Thus, the density \(\rho(t)\) is given by
\begin{equation}
   \rho(t) = - \frac{\partial }{\partial x} r(x,t)\big\vert_{x=0^+},
\label{rhor2}
\end{equation}
which proves Eq.~(\ref{rhor}).                  

\begin{figure}
\begin{center}
\leavevmode
\epsfxsize=6.0in
\epsffile{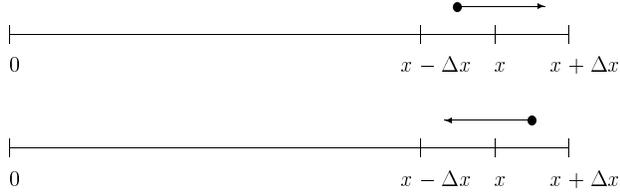}
\vspace{-5.75in}
\end{center}
\caption{Particle movements contributing to the
change in \(r(x,t)\). The effect in each case
depends on whether the number of particles in
\((0,x-\delx)\) is odd or even.}   
\label{moreintervals}
\end{figure}         

Next, we consider the intervals \((0,x-\delx)\),  \((x-\delx,x)\),
and \((x,x+\delx)\). Let ${\mathcal P}_{{\rm e}01}(x,\delx,t)$ be the
probability that there is an even number of particles in
\((0,x-\delx)\), no particle in \((x-\delx,x)\), and one particle
in \((x,x+\delx)\) at time \(t\). Let ${\mathcal P}_{{\rm e}00}(x,\delx,t)$,
${\mathcal P}_{{\rm e}10}(x,\delx,t)$, ${\mathcal P}_{{\rm e}11}(x,\delx,t)$,
${\mathcal P}_{{\rm o}00}(x,\delx,t)$, ${\mathcal P}_{{\rm o}01}(x,\delx,t)$,
${\mathcal P}_{{\rm o}10}(x,\delx,t)$, and ${\mathcal P}_{{\rm o}11}(x,\delx,t)$
be defined in the obvious way.  The appropriate intervals are shown in
Fig.~\ref{moreintervals}.
 
Because \(g(0,t)=0\) for all $t>0$, the probability that there are two
particles in \((x-\delx,x+\delx)\) is proportional to \(\delx^3\) as
\(\delx\to 0\). Thus,
\begin{eqnarray}
&& r(x + \delx,t) + r(x - \delx,t) - 2r(x,t) \nonumber \\ [12pt]
&&\qquad \qquad =
   {\mathcal P}_{{\rm e}10}(x,\delx,t)
    - {\mathcal P}_{{\rm o}10}(x,\delx,t)\nonumber\\[12pt]
&&\qquad \qquad - {\mathcal P}_{{\rm e}01}(x,\delx,t)
   + {\mathcal P}_{{\rm o}01}(x,\delx,t) + \Or{\delx^3}.
\label{rpp}
\end{eqnarray}              

We derive the contribution due to diffusion of particles to the
partial differential equations (\ref{drdt}) and (\ref{drdtb})
for the evolution of \(r(x,t)\) by
considering the probability that a particle at \(x-\delx\) at time
\(t\) diffuses out of the region \((0,x)\) before time \(t + \delt\),
and the probability that a particle at \(x+\delx\) at time \(t\)
diffuses out of the region \((0,x)\) before time \(t + \delt\).         

The probability that a particle, at \(x+\delx\) at time \(t\),
is in \((0,x)\) at time \(t+\delt\) is given by \cite{kands,expstep}
\begin{eqnarray}
   Q(\delx,\delt) = \left(4\pi D\right)^{-\frac12}
   \int_{\delx}^{\infty} \d x \; \exp\left(-x^2/4D\delt\right) &=&
   \frac12\erfc\left(\frac{\delx}{(4D\delt)^{\frac12}}\right)\nonumber\\
\label{ss}
\end{eqnarray}
 
Let
\begin{equation}
   R_{{\rm e}01}(x,\delx,t) = \frac{\partial}{\partial \delx}
   {\mathcal P}_{{\rm e}01}(x,\delx,t).
\label{redef}
\end{equation}
As \(\delx \to 0\), this quantity is the probability density for finding
a
particle at \(x+\delx\) at time \(t\), given that the
number of particles in \((0,x)\) is even. Similarly, let
\begin{equation}
   R_{{\rm e}10}(x,\delx,t) = \frac{\partial}{\partial \delx}
   {\mathcal P}_{{\rm e}10}(x,\delx,t),\qquad
   R_{{\rm o}01}(x,\delx,t) = \frac{\partial}{\partial \delx}
   {\mathcal P}_{{\rm o}01}(x,\delx,t),
\label{redefbis}
\end{equation}                   
and so on.

The time derivative of \(r(x,t)\)
\begin{equation}
   \frac{\partial}{\partial t}r(x,t) =
   \lim_{\delt\to 0}\frac1{\delt}(r(x,t+\delt)-r(x,t)),
\label{drdtdef}
\end{equation}
is found using Eq.~(\ref{rpp}), integrating over \(\delx\), and
taking the limit \(\delt\to0\)
\begin{eqnarray}
   r(x, t + \delt) - r(x,t)
   &=& 2\int_0^{\infty} \d \delx \; Q(\delx,\delt)
   \left(R_{{\rm e}10}(x,\delx,t) - R_{{\rm o}10}(x,\delx,t)\right)
   \nonumber\\ [12pt]
   &&\;\;  +2 \int_0^{\infty} \d \delx \;  Q(\delx,\delt)
   \left(- R_{{\rm e}01}(x,\delx,t) + R_{{\rm o}01}(x,\delx,t)\right)
   \nonumber\\[12pt]
   &=&2\int_0^{\infty} \d \delx \;  Q(\delx,\delt)
   \frac{\partial}{\partial\delx}\Big({\mathcal P}_{{\rm e}10}(x,\delx,t)
   - {\mathcal P}_{{\rm e}10}(x,\delx,t) \nonumber\\ [12pt] && \;\;
   - {\mathcal P}_{{\rm e}01}(x,\delx,t)
   + {\mathcal P}_{{\rm o}01}(x,\delx,t)\Big)\nonumber\\[12pt]&=&
   2\int_0^{\infty} \d \delx \; Q(\delx,\delt)\frac{\partial}{\partial\delx}
   \left(\frac{\partial^2}{\partial x^2}r(x,t)\delx^2 +
\Or{\delx^3}\right)
   \nonumber\\ [12pt]&=&
   \int_0^{\infty} \d \delx \; \erfc\left(\frac{\delx}{(4D\delt)^{\frac12}}\right)
   \left(\frac{\partial^2}{\partial x^2}r(x,t)\delx + \Or{\delx^2}\right)
   \nonumber\\[12pt]&=&2D\frac{\partial^2}{\partial x^2} r(x,t)\delt
    + \Or{\delt^2}.
\label{drdt2}
\end{eqnarray}        

\end{document}